\documentclass[twocolumn,showpacs,preprintnumbers,amsmath,amssymb，groupedaddress,superscriptaddress]{revtex4}
\usepackage{graphicx}
\usepackage{dcolumn}
\usepackage{multirow}
\usepackage{bm}
\usepackage{booktabs}
\usepackage{subfigure}

\begin{document}

\title{\emph{Ab initio} prediction of the mechanical properties of alloys: The case of Ni/Mn-doped ferromagnetic Fe}

\author{Guisheng Wang}
\email{guisheng@kth.se}
\affiliation{Applied Materials Physics, Department of Materials Science and Engineering, Royal Institute of Technology, Stockholm SE-100 44, Sweden}

\author{Stephan Sch\"onecker}
\email{stesch@kth.se}
\affiliation{Applied Materials Physics, Department of Materials Science and Engineering, Royal Institute of Technology, Stockholm SE-100 44, Sweden}

\author{Staffan Hertzman}
\affiliation{Department of Materials Science and Engineering, Outokumpu Stainless Research Foundation, Royal Institute of Technology, Stockholm SE-100 44, Sweden}

\author{Qing-Miao Hu}
\affiliation{Shenyang National Laboratory for Materials Science, Institute of Metal Research, Chinese Academy of Sciences, 72 Wenhua Road, Shenyang 110016, China}

\author{B\"orje Johansson}
\affiliation{Applied Materials Physics, Department of Materials Science and Engineering, Royal Institute of Technology, Stockholm SE-100 44, Sweden}
\affiliation{Department of Physics and Astronomy, Division of Materials Theory, Uppsala University, Box 516, SE-75121, Uppsala, Sweden}

\author{Se Kyun Kwon}
\affiliation{Graduate Institute of Ferrous Technology, Pohang University of Science and Technology, Pohang 790-784, Korea}
\author{Levente Vitos}
\affiliation{Applied Materials Physics, Department of Materials Science and Engineering, Royal Institute of Technology, Stockholm SE-100 44, Sweden}
\affiliation{Department of Physics and Astronomy, Division of Materials Theory, Uppsala University, Box 516, SE-75121, Uppsala, Sweden}
\affiliation{Research Institute for Solid State Physics and Optics, Wigner Research Center for Physics, P.O. Box 49, H-1525 Budapest, Hungary}

\date{\today}

\begin{abstract}
First-principles alloy theory, formulated within the exact muffin-tin orbitals method in combination with the coherent-potential approximation, is used to study the mechanical properties of ferromagnetic body-centered cubic (bcc) Fe$_{1-x}$M$_x$ alloys (M=Mn or Ni, $0\le x \le 0.1$). We consider several physical parameters accessible from  \emph{ab initio} calculations and their combinations in various phenomenological models to compare the effect of Mn and Ni on the properties of Fe. Alloying is found to slightly alter the lattice parameters and produce noticeable influence on elastic moduli. Both Mn and Ni decrease the surface energy and the unstable stacking fault energy associated with the $\{110\}$ surface facet and the $\{110\}\langle111\rangle$ slip system, respectively. Nickel is found to produce larger effect on the planar fault energies than Mn. The semi-empirical ductility criteria by Rice and Pugh consistently predict that Ni enhances the ductility of Fe but give contradictory results in the case of Mn doping. The origin of the discrepancy between the two criteria is discussed and an alternative measure of the ductile-brittle behavior based on the theoretical cleavage strength and single-crystal shear modulus $G\{110\}\langle111\rangle$ is proposed.
\end{abstract}


\maketitle

\section{Introduction}\label{Introduction}

Duplex stainless steels (DSS) have achieved great success in the development of stainless steels and have led to  numerous commercial applications. They consist of approximately equal amounts of ferrite and austenite. Though DSS successfully combined the merits of these two phases, the brittle cleavage problem in the ferrite component has attracted further consideration~\cite{Wessman2008}. In order to grasp the mechanism of ductility in DSS, in the present work we concentrate on the effect of Mn and Ni additions on the intrinsic properties of ferromagnetic Fe. Both alloying elements are fundamental constituents of DSS and they are primarily responsible for the stabilization of the austenite phase in the low-carbon alloys.

The mechanical properties of engineering alloys are determined by a series of mechanisms controlling the dislocation glide (such as elasticity, grain size, texture, grain boundary cohesion, solid solution and precipitation hardening). Describing such phenomena using merely \emph{ab initio} calculations is not feasible. On the other hand, the mechanical properties are closely related to several material parameters which are within reach of such theoretical tools. In particular, the fracture work depends on the surface energy, elastic energy, and plastic energy~\cite{Pugh1954}. Following the relationship between the crack and the surface energy ($\gamma_s$), and the plastic deformation and the unstable stacking fault (USF) energy ($\gamma_u$), Rice \cite{Rice1992} proposed that the brittle-ductile behavior can be effectively classified by the $\gamma_s/\gamma_u$ ratio. In body-centered cubic (bcc) system, plastic deformation shows various kinds of complexity due to its intricate dislocation core structure, multiple slip systems and the cross slip between them. According to the Peierls-Narbarro model, the shear stress necessary for moving a dislocation in the lattice can be estimated as $\tau_p=2G/(1-\nu)\exp[-2a/(b(1-\nu))]$, where $G$ is the shear modulus, and $\nu$ the Poisson ratio. Here $a$ represents the characteristic interlayer distance, and $b$ is the Burgers vector. For a fixed structure, $a/b$ depends only on the slip system. Accordingly, the greatest $a$ and the shortest $b$ give the lowest resistant stress, which corresponds to the most likely slip system. Usually, such slip systems have the closest packed atomic plane and the closest packed slip direction. In the bcc lattice, this is the $\{110\}\langle111\rangle$ slip system, \emph{i.e.}, slip occurs along the $\langle111\rangle$ direction on the $\{110\}$ planes. A recent molecular dynamics study \cite{Kumar2013} showed that other plane families such as $\{112\}$ and $\{123\}$ can be also efficient in bcc Fe.

Unfortunately, both the USF energy and the surface energy for a particular crystallographic facet are extremely difficult to verify from experiments, and thus today the limited available data are exclusively based on first-principles quantum-mechanical calculations. However, long before the era of ``\emph{ab initio} materials science'', Pugh summarized dozens of experimental data and developed a simple phenomenological model~\cite{Pugh1954} which correlates the fracture properties with some easily accessible elastic parameters. According to this model, the ductility of materials can be characterized by the ratio between the bulk modulus $B$ and the shear modulus $G$. Ductile materials usually possess $B/G>\chi_{\rm P}\equiv 1.75$, whereas brittle materials have typically $B/G<\chi_{\rm P}$. The Pugh ratio is often used in modern computational materials science to assess and predict trends at relatively low calculation cost.

One can easily establish a qualitative connection between the Pugh criterion and that of Rice. The surface energy is known to correlate with the cohesive energy \cite{Vitos1994,Vitos1997,Kollar2000,Kwon2005,PRL69}, which in turn is to a large degree proportional to the bulk modulus since larger cohesive energy requires larger curvature of the binding energy curve. The USF energy represents the energy barrier upon shifting one half of the crystal over the other half. The slope of the so defined $\gamma$-surface is expected to be proportional to $\gamma_u$ (maximum of the curve) and also to the Peierls-Narbarro stress. Assuming a constant Poisson ratio, for a fixed lattice the latter is given merely by the shear modulus. Hence, at least on a qualitative level one may expect that $\gamma_s/\gamma_u$ and $B/G$ should follow similar trends, \emph{e.g.}, upon alloying. The above correlations will be carefully investigated in the present work in the case of Fe-Mn and Fe-Ni alloys.

For an isotropic polycrystalline material the two elastic moduli $B$ and $G$ determine the Poisson ratio $\nu = (3B/G-2)/(6B/G+2)$. Consequently, the Pugh criterion imposes the constraint $\nu > (3\chi_{\rm P}-2)/(6\chi_{\rm P}+2) \approx 0.26$ for the Poisson ratio of ductile materials. The polycrystalline bulk and shear moduli are derived from single-crystal elastic constants by Hill averaging of the Voigt and Reuss bounds \cite{Vitos-book}. In the case of cubic crystals, these two bounds are related to Pugh's ratio by
\begin{eqnarray}\label{eq0}
\frac{B}{G} = \left\{\begin{array}{cc}5\left(\frac{2}{3}Z^{-1}+\mu\right)(2Z^{-1}+3)^{-1}\;\;\mbox{(Voigt)}\\
\frac{1}{5}\left(\frac{2}{3}Z^{-1}+\mu\right)(3+2Z)\;\;\mbox{(Reuss)}
\end{array}\right.,
\end{eqnarray}
where $Z\equiv 2C_{44}/(C_{11}-C_{12})$ is the Zener anisotropy ratio. Hence, for both limiting cases, the Pugh ratio increases linearly with $\mu\equiv C_{12}/C_{44}$. For isotropic crystals $Z=1$ and thus the Pugh criterion for ductility requires $\mu\gtrsim 1.08$. For weakly anisotropic lattices ($Z\sim 2$, which is the case of the present alloys), the above condition modifies to $\eta\gtrsim 0.92$ or $\eta\gtrsim 1.07$ corresponding to the Reuss or Voigt value, respectively. For this anisotropy level, the Hill average requires $\mu\gtrsim 0.99$ for ductile materials. The above defined $\mu$ parameter is closely connected to the Cauchy pressure $(C_{12}-C_{44})=C_{44}(\mu-1)$. Negative Cauchy pressure (\emph{i.e.}, $\mu<1$) implies angular character in the bonding and thus characterizes covalent crystals while positive Cauchy pressure ($\mu>1$) is typical for metallic crystals \cite{Pettifor1992}. In other words, if the resistance of the crystal to shear ($C_{44}$) is larger than the resistance to volume change ($C_{12}$), the system shows covalent features and vice versa. This condition based on the Cauchy pressure is in line with the one formulated by Pugh.

Compared to the criteria based on Pugh ratio and Cauchy pressure, the condition developed by Rice is expected to give more physical insight into the mechanism of plastic deformation. However, the application of Rice condition based on planar defect energies is related to the specific deformation mode and crystal orientation. In principle, one should analyze all possible cases with appropriate statistical weights. Except simple cases, such as the face-centered cubic (fcc) lattice, this makes the Rice approach rather cumbersome. In turn, the Pugh rule was established from a series of experimental data and is based on simple physical quantities. Although many assumptions are involved in this empirical approach, it has been confirmed to be practical in a large number of former scientific studies.

In the present work, we focus on the alloying effect of Ni and Mn on ferrite Fe-based dilute alloys. In order to study the possible mechanisms behind the well-known facet cleavage problem in DSS alloys, which to a large extent limits their engineering applications, here we carry out systematic electronic structure calculations to find the single-crystal and polycrystalline elastic parameters and the surface and USF energies as a function of chemical composition.
We consider the $\{110\}\langle111\rangle$ slip system due to its special role in the plastic deformation of the bcc lattice. Accordingly, we compute the surface energy for the $\{110\}$ crystallographic facet and the USF energy for the $\{110\}$ plane along the $\langle111\rangle$ slip direction. The theoretical predictions are used to assess various ductility parameters discussed above.

The rest of the paper is divided into three main sections and conclusions. Section~\ref{tools} gives a brief description of the employed theoretical tools, an introduction to the theory of elasticity and planar defects, and lists the important numerical details. The results are presented and discussed in Sections~\ref{result} and \ref{discussion}, respectively.

\section{Theoretical tools}\label{tools}

\subsection{Total energy method}\label{method}
The total energy calculations were performed using the exact muffin-tin orbitals (EMTO) method~\cite{Vitos2000, Vitos2001, Vitos-PRL, Vitos-book}. This density functional theory (DFT) solver is similar to the screened Korringa-Kohn-Rostoker method, where the exact Kohn-Sham potential is represented by large overlapping potential spheres. Inside these spheres the potential is spherically symmetric and constant between the spheres. It was shown~\cite{Andersen1998, Zwierzycki2009, Vitos-book} that using overlapping spheres gives a better representation of the full-potential as compared to the traditional muffin-tin or atomic-sphere approximations. Within the EMTO method, the compositional disorder is treated using the coherent-potential approximation (CPA)~\cite{Soven1967, Gyorffy1972} and the total energy is computed via the full-charge density technique~\cite{Kollar2000, Vitos1997, Kollar1997}. The EMTO-CPA method has been involved in many successful applications focusing on the thermo-physical properties of alloys and compounds~\cite{Delczeg2011,GJ,Landa2006,Magyari2001,Huang2006,Kollar2003,Magyari2004,Hu2009,Zhang2009,Zander2007,Vitos2006}, surface energies~\cite{Schonecker2013,Ropo2005} and stacking-fault energies~\cite{Li2014,Lu2011,Lu2012}. Here we would like to make a brief comment on CPA. Due to the single-site nature, CPA cannot account for the effect of local concentration fluctuations, short-range order effect and local lattice relaxation. This means that modeling the alloy by CPA one implicitly assumes a completely homogeneous random solid solution with a rigid underlying crystal lattice. On the other hand, CPA can be used to describe fluctuations within the supercell, e.g. near the unstable stacking fault or surface, by using variable concentrations for each atomic layer. This feature is used here as well when studying the segregation effects (see Section II.C). In the case of Fe-Ni and Fe-Mn alloys, embedding a single impurity atom into the mean-field (CPA) Green function means that we omit the inter-dependence between magnetic state and the local environment, which in turn could lead to specific local relaxation effects. In particular, within the present approximation the magnetic state of Mn is determined by the coherent Green function (representing the average effect of all alloying elements) rather than by the actual nearest neighbor atoms. Nevertheless, such local effects are expected to be relatively small when the impurity level is below $\sim10$\% (i.e. all impurities are mostly surrounded by Fe).

The self-consistent EMTO calculations were performed within the generalized-gradient approximation proposed by Perdew, Burke, and Ernzerhof (PBE)~\cite{Perdew1996}. The performance of this approximation was verified for the Fe-based systems in many former investigations~\cite{Zhang2010,Asker2009,Pitkanen2009}. The one-electron equations were solved within the soft-core scheme and using the scalar-relativistic approximation. The Green's function was calculated for 16 complex energy points distributed exponentially on a semicircular contour including states within 1 Ry below the Fermi level. In the basis set, we included \emph{s, p, d}, and \emph{f} orbitals and for the one-center expansion of the full-charge density an orbital cut-off $l_{max}=8$ was used. The total energy was evaluated by the shape function technique with cut-off $l_{max}^{shape} = 30$~\cite{Vitos-book}. For the undistorted bcc structure, we found that a homogeneous $k$-mesh of $37\times37\times37$ ensured the required accuracy. For the elastic constant calculations, we used about $\sim30000-31000$ uniformly distributed $k$ points in the Brillouin zones of the orthorhombic and monoclinic structures. The potential sphere radii for the alloy components were fixed to the average Wigner-Seitz radius. For the surface energy and USF energy calculations, the $k$-mesh was set to $9\times23\times2$ in the the Brillouin zone of the base-centered orthorhombic super-cell.

The impurity problem was solved within the single-site CPA approximation, and hence the Coulomb system of a particular alloy component $i$ may contain a non-zero net charge. Here the effect of charge misfit was taken into account using the screened impurity model (SIM) \cite{Korzhavyi1995,Ruban2002}. According to that, the additional shift in the one-electron potential and the corresponding correction to the total energy are controlled by the dimensionless screening parameters $\alpha_i$ and $\beta$, respectively. The parameters $\alpha_i$ are usually determined from the average net charges and electrostatic potentials of the alloy components obtained in regular supercell calculations \cite{Ruban2002}. The second dimensionless parameter ${\beta}$ is determined from the condition that the total energy calculated within the CPA should match the total energy of the alloy obtained using the supercell technique. For most of the alloys, the suggested optimal values of ${\alpha_i}$ and ${\beta}$ are $\sim0.6$ and $\sim1.2$, respectively \cite{Korzhavyi1995,Ruban2002}. Often $\alpha_i$ for different alloy components are assumed to the same and $\beta=1$ is used.

In the present work, considering the volume sensitivity of $\alpha_i$, a dynamic SIM \cite{Guisheng2013} was applied for the Fe-Ni system. According to that, the optimal $\alpha_i$ varies as $1.099, 1.130, 1.160, 1.179, 1.186, 1.182, 1.179, 1.172, 1.163$ for $9$ lattice parameters $a$ ranging from $\sim 2.805$ to $\sim 2.891$ \AA\  (with interval of $\sim 0.011$ \AA). The second SIM parameter ${\beta}$ was fixed to 1. For Fe-Mn, the SIM parameters show rather weak volume dependence, so ${\alpha_i}$ and $\alpha\equiv\beta\alpha_i$ were chosen to be $0.79$ and $0.90$, respectively. The above figures for the SIM parameters were obtained for a specific concentration (6.25 \% Ni or Mn in Fe) and thus they are strictly valid only for that particular disordered system. However, in this study we assumed the same values for all binaries, \emph{i.e.}, the concentration dependence of the SIM parameters was omitted.

\subsection{Elastic properties}\label{elastic}

The elastic properties of a single-crystal are described by the elements of the elasticity tensor. In a cubic lattice, there are three independent elastic constants: $C_{11}$, $C_{12}$ and $C_{44}$. The tetragonal shear elastic constant ($C^{\prime}$) and the bulk modulus ($B$) are connected to the single-crystal elastic constants through $B=(C_{11}+2C_{12})/3$ and $C^{\prime}=(C_{11}-C_{12})/2$.

The adiabatic elastic constants are defined as the second order derivatives of the energy ($E$) with respect to strain. Accordingly, the most straightforward way to obtain the elastic parameters is to strain the lattice and evaluate the total energy change as a function of lattice distortion. In practice, the bulk modulus is derived from the equation of state, obtained by fitting the total energy data calculated for a series of volumes ($V$) or cubic lattice constants ($a$) by a Morse-type function~\cite{Moruzzi1988}. Since we are interested in equilibrium (zero pressure) elastic parameters, the fitting should involve lattice points nearby the equilibrium. It is known that bcc Fe undergoes a weak magnetic transition at slightly enlarged lattice parameters ($a \approx 2.9\,\textrm{\AA}$), which should be excluded from the fitting interval to avoid undesirable scatter of the computed bulk modulus~\cite{Zhang2010a}. Accordingly, the present Morse-type fit was limited to lattice parameters smaller than $2.9\,\textrm{\AA}$. The same interval was applied to all considered binary alloys to minimize the numerical noise due to the volume mesh.

Since the total energy depends on volume much more strongly than on small lattice strains, volume-conserving distortion are usually more appropriate to calculate $C^{\prime}$ and $C_{44}$. Here, we employed the following orthorhombic ($D_o$) and monoclinic ($D_m$) deformations
\begin{eqnarray}\label{eq1}
D_o&=&\left(
  \begin{array}{ccc}
    1+\delta_o & 0 & 0 \\
    0 & 1-\delta_o & 0 \\
    0 & 0 & \frac1{1-{\delta^2_o}} \\
  \end{array}
\right)
\;\;\mbox{and}\;\; \nonumber \\
D_m&=&\left(
      \begin{array}{ccc}
        1 & \delta_m & 0 \\
        \delta_m & 1 & 0 \\
        0 & 0 & \frac{1}{1-{\delta^2_m}} \\
      \end{array}
    \right)\nonumber,
    \end{eqnarray}
where $\delta$ denotes the distortion parameter. For both lattice deformations, six different distortions $\delta = 0,0.01, 0.02, ..., 0.05$ were used. These deformations lead to total energy changes $\Delta E(\delta_{o}) = 2VC^{\prime} \delta_o^2 + \mathcal{O}(\delta_o^4)$ and $\Delta E(\delta_{m}) = 2VC_{44} \delta_m^2 + \mathcal{O}(\delta_m^4)$, respectively, where $\mathcal{O}$ stands for the neglected terms.

The polycrystalline shear modulus was obtained from the single-crystalline elastic constants according to the Hill average~\cite{Hill1952}, $G=(G_V+G_R)/2$, of the Voigt, $G_V=(C_{11}-C_{12}+3C_{44})/5$, and Reuss, $G_R=5(C_{11}-C_{12})C_{44}/[4C_{44}+3(C_{11}-C_{12})]$, bounds.

\begin{figure}
	\centering
	\subfigure[]{\includegraphics[scale=0.4]{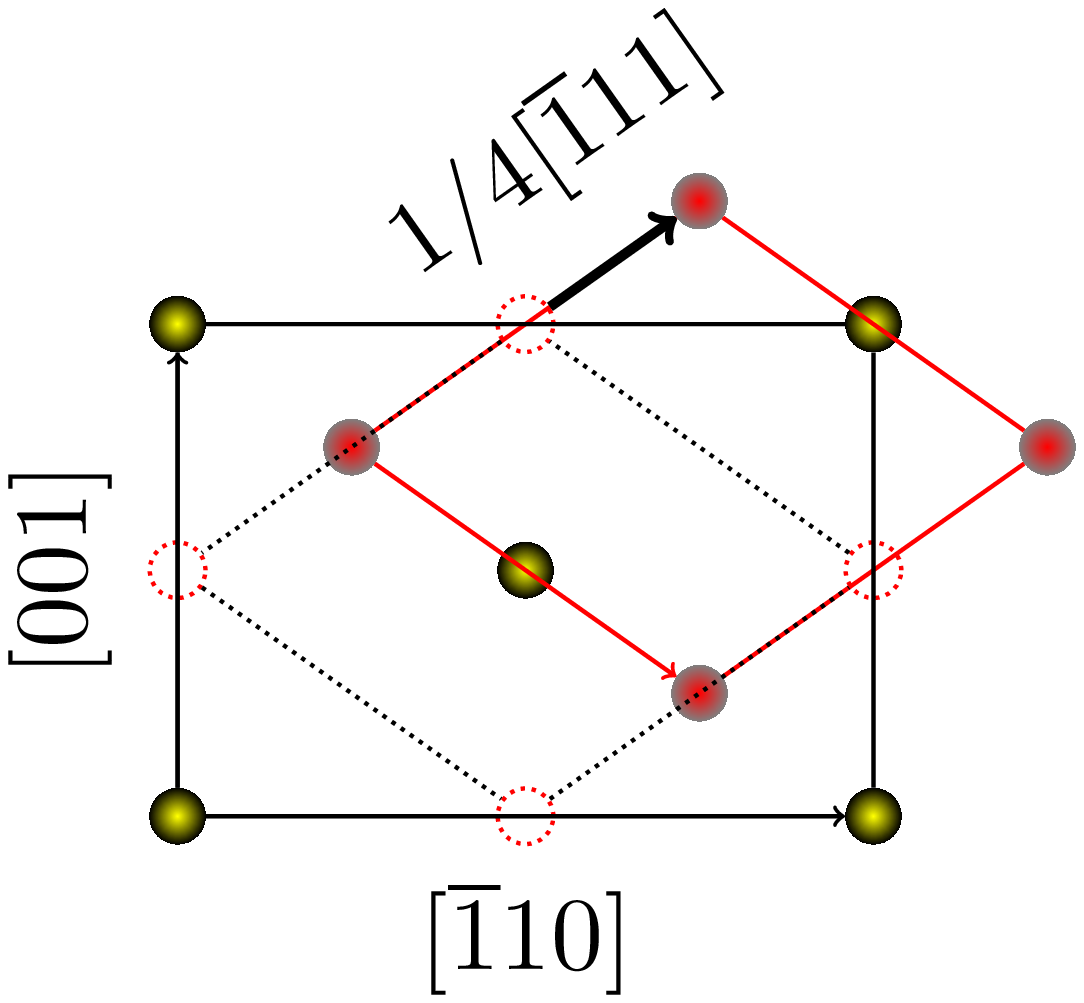}}\hspace{0.4in}
	\subfigure[]{\includegraphics[scale=0.4]{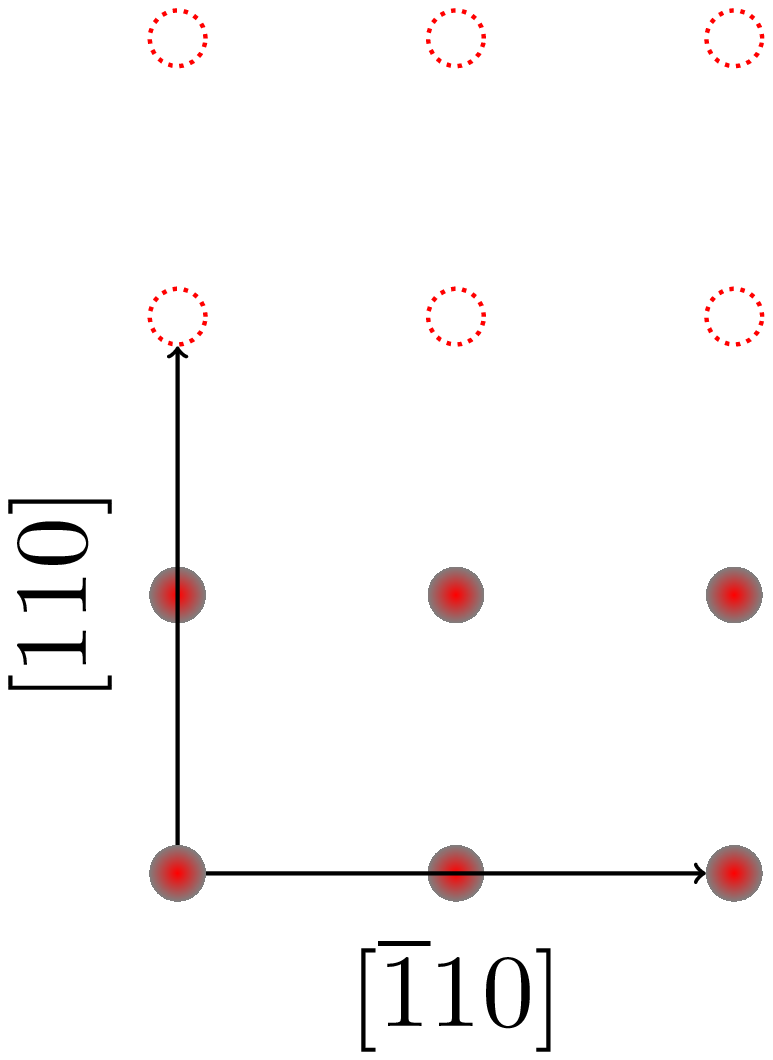}}
	\caption{(Color online) Schematics of the model structures used to compute the USF energy (panel a) and the surface energy (panel b). See text for details. }\label{fig1}
\end{figure}

\subsection{Surface and unstable stacking fault energy}

The schematics of the USF and surface model are illustrated in Fig.~\ref{fig1}. For USF (panel a), we show a top view along the $[110]$ direction of the two atomic layers next to the stacking fault. The slip direction for the second (filled open circles) atomic layer is also shown. For the surface model (panel b), we show the side view along the $[001]$ direction of the two surface atomic layers (filled circles) and the two empty layers (empty circles) mimicking the vacuum.

The surface energy ($\gamma_s$) of chemically homogeneous alloys (without segregation) was extracted from total energy calculations for slabs following the method proposed by Fiorentini and Methfessel~\cite{Fiorentini1996}. No reference to separately computed bulk total energies is made in this approach. Instead, the reference bulk energy per layer is deduced from a linear fit to the slab total energy versus the slab thickness. Slabs with 8, 10, and 12 atomic layers separated by 6 vacuum layers were used. According to Punkkinen \emph{et al.}~\cite{Punkkinen2011} the surface relaxation effects are small for the $\{110\}$ surface facet of Fe and have an insignificant effect on the magnitude of the surface energy. Our calculations confirmed this finding, and therefore here we omitted the effect of relaxation in all surface energy calculations.

The USF energy ($\gamma_u$) was obtained in the following way. First, the total energies of 16, 20, and 24 layers thick supercells (each containing two USF) were calculated, and the bulk energy per layer was extracted from a linear fit to the energy versus thickness.  Because the USF seriously disturbs the atomic arrangement in the vicinity of the stacking fault plane, structural relaxations perpendicular to the fault plane have to be taken into account. We considered the relaxation effect for pure bcc Fe and found that the first inter-layer spacing at the USF increases by approximately $4 \%$. For the Fe-based alloys, we kept the Mn/Ni content below $10 \%$, which is expected to have a small additional effect on the layer relaxation. Therefore, we used the structural relaxation obtained for pure Fe for all binaries considered here.

We also considered the effect of segregation on the surface energy and the USF energy. To this end, we varied the composition of the atomic layers next to the stacking fault or at the surface. Unfortunately, the above described method to compute the formation energies based entirely on the slabs/supercells with different sizes is not applicable when the chemical concentration profile is no longer homogeneous. Instead, all segregation studies were performed for a slab/supercell with fixed size and the impact of segregation was interpreted with respect to the homogeneous slab/supercell with the same size. In that way, possible numerical errors associated with the reference system were kept at minimal level. In particular, the effect of segregation on the USF energy was derived by subtracting the energy of the supercell without USF from the energy of the supercell with USF, both with segregation on the slip planes. The surface energy for the chemically inhomogeneous systems was computed following the method proposed in Ref.~\cite{ruban1999surface} based on the grand canonical ensemble and involving the effective chemical potentials.

\section{Results: intrinsic material parameters}\label{result}

\subsection{Parameters of ferromagnetic bcc Fe}

We assess the accuracy of the EMTO method for the present systems by comparing the results obtained for pure Fe with those derived from previous full-potential calculations and the available experimental data. Results are listed in Table~\ref{tabI}. The present lattice parameter $a$ is consistent with those from Refs.~\cite{Caspersen2004,Guo2000}. All theoretically predicted equilibrium lattice parameter are, however, slightly smaller than the experimental value, $2.866$\,\AA~\cite{Rayne1961}, which is due to the weak PBE over binding. Theory reproduces the experimental magnetic moment accurately.

\begin{table*}
\caption{Theoretical and experimental lattice parameter $a$ (in units of \AA), magnetic moment $\mu$ (in $\mu_B$), single-crystal elastic constants (in GPa), surface energy $\gamma_s$ (in J/m$^2$) and USF energy $\gamma_u$ (in J/m$^2$) for ferromagnetic bcc Fe. The quoted theoretical methods are EMTO: present results; PAW: full-potential projector augmented wave method; PP: ultrasoft non-norm-conserving pseudopotential; FLAPW: all-electron full-potential linear augmented plane wave method; FPLMTO: full-potential linear muffin-tin orbitals method. The low-temperature (4 K) experimental elastic constants are listed for comparison, as well as the semi-empirical surface energy.}\label{tabI}
\begin{ruledtabular}
\begin{tabular}{lccccccccc}
Method&  $a$  & $\mu$ & $C_{11}$ & $C_{12}$ & $C_{44}$ & $C^\prime$ & $B$ & $\gamma_s$ & $\gamma_u$ \\
EMTO  & 2.838  & 2.21  & 288.9 & 133.7 & 100.0 & 77.6 & 185.4 & 2.47 & 1.08\\
PAW   & 2.838\cite{Caspersen2004}  &  2.21\cite{Caspersen2004}  & 271\cite{Caspersen2004}   & 145\cite{Caspersen2004}   & 101\cite{Caspersen2004}&63\cite{Caspersen2004} & 172\cite{Caspersen2004}&2.50\cite{Punkkinen2011} & 0.98\cite{Mori2009}\\
PP & 2.848\cite{Vovcadlo1997}  & 2.25\cite{Vovcadlo1997}  & 289\cite{Vovcadlo1997}   & 118\cite{Vovcadlo1997}  &115\cite{Vovcadlo1997} & 85.5\cite{Vovcadlo1997}&176\cite{Vovcadlo1997}&  \\
FLAPW & 2.84\cite{Guo2000} & 2.17\cite{Guo2000} & 279\cite{Guo2000} & 140\cite{Guo2000} & 99\cite{Guo2000} & 69\cite{Guo2000} & 186\cite{Guo2000} & &\\
FPLMTO & 2.812\cite{Sha2006}  &       & 303\cite{Sha2006}   & 150\cite{Sha2006}  &126\cite{Sha2006} & 76.5\cite{Sha2006}&201\cite{Sha2006}&  \\
Expt. & 2.866\cite{Rayne1961}  & 2.22\cite{Rayne1961}  & 243.1\cite{Rayne1961} &138.1\cite{Rayne1961} & 121.9\cite{Rayne1961} & 52.5\cite{Rayne1961} & 173.1\cite{Rayne1961} & 2.41 \cite{Tyson1977}& \\
\end{tabular}
\end{ruledtabular}
\end{table*}

In Table~\ref{tabI}, the present single-crystal elastic constants and the bulk modulus of bcc Fe are compared with data from literature. The EMTO results are consistent with the other theoretical elastic constants, the deviations being typical to the errors associated with such calculations. However, compared to the experimental data, EMTO is found to overestimate $C_{11}$ by about $\sim 19 \%$, and underestimates $C_{44}$ by $\sim 18 \%$. The larger $C_{11}$ (and the larger bulk modulus) can be partly accounted for by the underestimated lattice parameter~\cite{Zhang2010}. We notice that the excellent agreement in Ref.~\cite{Caspersen2004} between the theoretical and experimental bulk moduli is somewhat suspicious since the quoted single-crystal elastic constants results in $187$ GPa for the bulk modulus (compared to $172$ GPa reported in Ref.~\cite{Caspersen2004}), which is close to the present finding and also to that from Ref.~\cite{Guo2000}. The pseudopotential~\cite{Vovcadlo1997} equilibrium lattice parameter is the largest among the theoretical predictions listed in the table, which might explain why the pseudopotential bulk modulus is surprisingly close to the experimental value.

The EMTO surface energy is in close agreement with the full-potential result~\cite{Punkkinen2011} and also with the semi-empirical value valid for an ``average'' surface facet~\cite{Tyson1977}. The present USF energy ($1.08$\,J/m$^2$) reproduces the full-potential result from Ref.~\cite{Mori2009} within $\sim 10\%$, which is well acceptable taking account that the present method is based on muffin-tin and full-charge density approximations. We recall that today it is not yet possible to measure the USF energy.

\subsection{Equilibrium volumes and magnetic moments of Fe-Ni and Fe-Mn alloys}

\begin{figure}[b]
\includegraphics[scale=0.3]{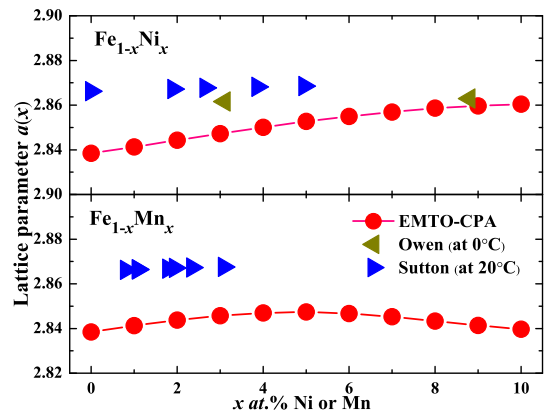}
\caption{(Color online) Lattice parameters (in units of \AA) of bcc Fe$_{1-x}$Mn$_x$ and Fe$_{1-x}$Ni$_x$ alloys as a function of composition. For comparison, the measured values by Sutton~\cite{Sutton1955} and Owen~\cite{Owen1937} are also included.}\label{fig2}
\end{figure}

According to the equilibrium phase diagrams, in ferromagnetic bcc Fe the maximum solubility of Mn is $\sim 3\%$ and that of Ni is $\sim 5.5\%$. Nevertheless, in duplex stainless steels, solid solutions with substantially larger amount of Mn and Ni may appear as a result of the delicate balance between volume and surface effects. Hence, in the present theoretical study we go well beyond the experimental solubility limits and discuss the trends up to $10\%$ solute atoms in the Fe matrix. The calculated equilibrium lattice parameters for bcc binary Fe$_{1-x}$Mn$_x$ and Fe$_{1-x}$Ni$_x$ alloys are shown in Fig.~\ref{fig2} for $0\le x \le 0.1$. We find that for both binaries the lattice parameters show a weak dependence on the amount of alloying addition $x$. Our results, especially the alloying trends, are in reasonable agreement with the available experimental data~\cite{Sutton1955,Owen1937}.

The lattice parameter of bcc Fe increases with Mn addition as $x$ increases from $0$ to $0.05$. When Fe is alloyed with more than 5\,\% Mn, the lattice parameter shows a small negative slope as a function of $x$.
Due to the limited solubility of Mn in bcc Fe, the experimental data is restricted below $\sim 3 \%$ Mn.
In contrast to Mn, the lattice parameter of Fe$_{1-x}$Ni$_x$ monotonically increases with the Ni amount.
Sutton \emph{et al.}~\cite{Sutton1955} reported that the lattice parameter of Fe$_{1-x}$Ni$_x$ is a linear function of $x$ up to $\sim 5 \%$ Ni with small slope and remains constant at larger concentrations. This trend is reproduced by the present theory, although the switch from a weak positive slope to a constant value seems to be shifted to larger concentrations (Fig.~\ref{fig2}).

The present theory slightly overestimates the observed volume expansion due to alloying. This can have both theoretical and experimental origin. From the side of theory, the exchange-correlation functional (PBE) may produce a noticeable effect since it performs differently for the present metals. For instance, while PBE underestimates the equilibrium lattice parameter of bcc Fe by $\sim 1.0\%$, it slightly overestimates that of fcc Ni ($\sim 0.3\%$) \cite{Ropo2008,Punkkinen2011}. This error results in $\sim 0.004$ \AA\; ``extra'' increase of the lattice parameter at $10 \%$ Ni compared to the ideal case when the DFT error stays at the same level for different elements. A careful inspection of Fig. \ref{fig2} shows that the above error accounts for about 30\% of the deviation obtained between the theoretical and experimental slopes of $a(x)$ for the Fe-Ni system. Another possible source of error is the neglect of all thermal effects in the present theory. From the experimental point of view, not all measured data points are relevant for the present comparison. That is because both Ni and Mn are fcc stabilizer, and thus a perfect bcc single-phase is hard to attain. Moreover, in the process of preparing the samples, different quenching temperatures and various annealing methods lead to subtle changes in the measured lattice parameter.

The overall small influence of Mn/Ni alloying addition on the lattice parameter of Fe can be explained from the atomic volume and the magnetic moment. The atomic volumes of Mn and Ni are very close to that of Fe. Hence, based on the linear rule of mixing, small amounts of Mn and Ni should not affect the volume of Fe to a large extent. On the other hand, the observed small increment can be ascribed to the complex magnetic interaction between the solute atoms and ferromagnetic Fe matrix.

\begin{figure}
\includegraphics[scale=0.3]{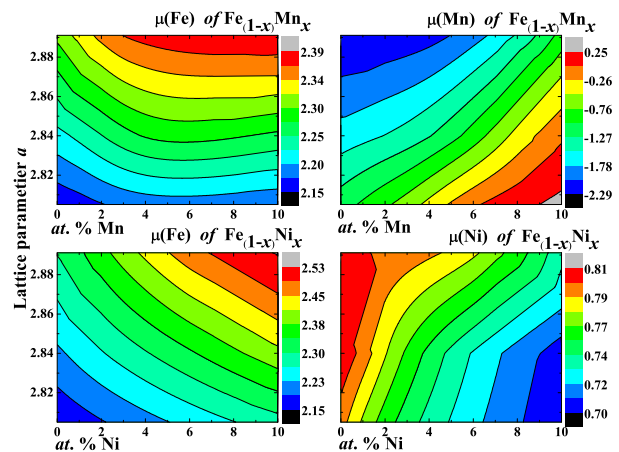}
\caption{(Color online) The map of the local magnetic moments $\mu(x)$ (in units of $\mu_B$) as a function of chemical composition and lattice parameter $a$ (in units of \AA) for Fe and Mn in Fe$_{1-x}$Mn$_x$ (upper panels) and Fe and Ni in Fe$_{1-x}$Ni$_x$ (lower panels).}\label{fig3}
\end{figure}

The local magnetic moments in Fe-Mn and Fe-Ni are displayed in Fig.~\ref{fig3} as a function of composition and lattice parameter. These results were extracted from the CPA calculations and the local magnetic moments represent the magnetic moment density integrated within the corresponding Wigner-Seitz cells. Around the equilibrium lattice parameters ($\sim 2.84-2.85$ \AA), the Mn moments are antiparallel to those of Fe. The magnitude of Mn local magnetic moment $\mu$(Mn) decreases from $\sim 1.7 \mu_{\rm B}$ to about zero as the amount of Mn increases from zero to $10\%$ (Fig. \ref{fig3}, upper right panel). This trend is consistent with the previous findings, namely that above $\sim 10 \%$ Mn, the coupling between Fe and Mn becomes ferromagnetic \cite{Kulikov1997}. At the same time, the magnetic moment of Fe $\mu$(Fe) is increased slightly with Mn addition up to $\sim 5\%$ Mn, above which a very weak decrease of $\mu$(Fe) can be observed (Fig. \ref{fig3}, upper left panel). The enhanced $\mu$(Fe) expands the volume as a result of the positive excess magnetic pressure \cite{Punkkinen2011a}. For $x\gtrsim 0.05$, the small negative slope of $\mu$(Fe) leads to vanishing excess magnetic pressure and thus to shrinking volume, in line with Fig. \ref{fig2}. On the other hand, the Ni moments $\mu$(Ni) always couple parallel with the Fe moments (Fig. \ref{fig3}, lower right panel) and Ni addition increases $\mu$(Fe) (Fig. \ref{fig3}, lower left panel). This in turn increases the excess magnetic pressure and yields monotonously expanding volume with $x$. The electronic structure origin of the alloying-induced enhancement for the Fe magnetic moment for the present binaries is discussed in Ref.~\cite{xiaoqing2014}.

In order to gain more insight into the magnetic coupling between Mn and the bcc Fe host, we constructed a $2\times2\times2$ supercell in terms of the conventional bcc unit cell (16-atom supercell) containing one or two Mn impurity atoms, which correspond to $6.25 \%$ Mn and $12.5 \%$ Mn, respectively. In the case of one isolated impurity atom, Mn couples antiferromagnetically to Fe and the average magnetic moments of Fe and Mn are almost identical to those from the corresponding CPA calculation. In the case of $12.5 \%$ Mn in the supercell, both Mn moments align parallel with respect to the surrounding Fe host irrespective of their relative positions in the supercell. However, the magnitude of their moments scales with the distance between them. If the two Mn atoms are located at nearest neighbor sites, their local magnetic moments are strongly reduced which may be interpreted as a result of competing magnetic interactions between Mn-Mn and Mn-Fe. At the same time, the local magnetic moments of the nearest Fe atoms remain almost at the level of pure bcc Fe, similar to the CPA result. These tests, based on supercell calculations, confirm that CPA accurately mimics the behavior of the magnetic moments versus concentration in the dilute limit. 
\subsection{Elastic parameters}

\begin{figure}[t]
\includegraphics[scale=0.3]{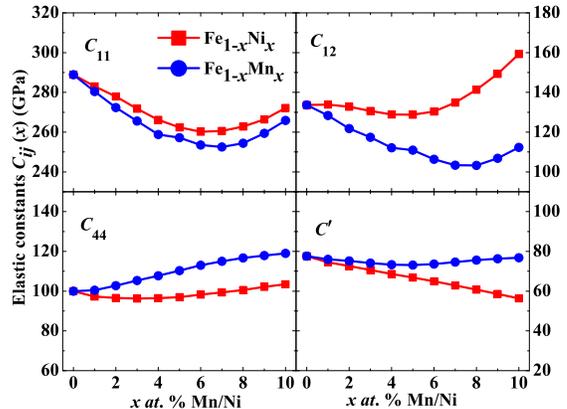}
\caption{(Color online) Single-crystal elastic constants of bcc Fe$_{1-x}$Mn$_x$ and Fe$_{1-x}$Ni$_x$ alloys as a function of composition.}\label{fig4}
\end{figure}

\begin{figure}[t]
\includegraphics[scale=0.3]{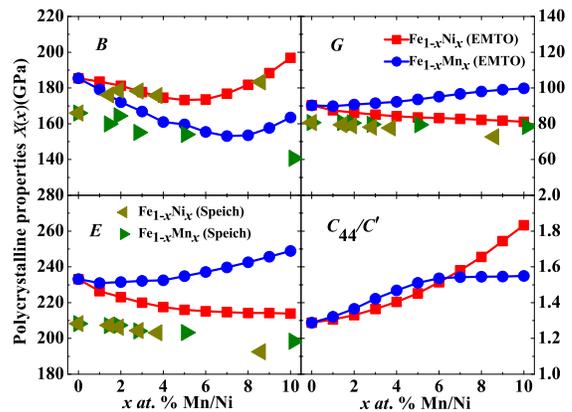}
\caption{(Color online) Polycrystalline elastic moduli of bcc Fe$_{1-x}$Mn$_x$ and Fe$_{1-x}$Ni$_x$ alloys as a function of composition. The experimental data are from Ref.~\cite{Speich1972}.}\label{fig5}
\end{figure}

The theoretical single-crystal elastic constants and polycrystalline elastic moduli of the Fe-Ni and Fe-Mn alloys are plotted in Figs.~\ref{fig4} and \ref{fig5} as a function of composition and the corresponding data are listed in Table~\ref{tabII}. The results indicate rather impressive alloying effects of Mn/Ni on the elastic parameters. We find that Mn and Ni produce similar effects on $C_{11}$ of Fe. Namely, $C_{11}$ decreases with $x$ below $\sim 7\%$ alloying addition, and then increases at larger concentrations. But Mn and Ni give different effects on $C_{12}$. Below $\sim 6\%$ Ni, $C_{12}$ keeps constant with Ni content, and then strongly increases with further Ni addition. On the other hand, $C_{12}$ drops from $133.7$ to $103.2$ GPa as the Mn concentration increases from zero to $\sim 8\%$. At larger concentrations, the effect of Mn on $C_{12}$ changes sign. The peculiar concentration dependencies obeyed by $C_{11}$ and $C_{12}$ originate mainly from the non-linear trend of the corresponding bulk modulus shown in Fig.~\ref{fig5}.

The two cubic shear elastic constants also exhibit complex trends. $C^{\prime}$ decreases with Ni addition, which means that Ni decreases the mechanical stability of the bcc lattice. At the same time, Mn is found to have very small impact on the tetragonal shear elastic constant. $C_{44}$ increases with Mn addition, but remains nearly constant upon alloying Fe with Ni. The present trends for the elastic constants are very close to those reported by Zhang \emph{et al.} \cite{Zhang2010} using the same total energy method. The small differences seen at larger concentrations, especially in the case of the bulk modulus, are due to the different numerical parameters used in these two works.

The theoretical polycrystalline elastic moduli $B$, $E$, and $G$ (Fig.~\ref{fig5}) reproduce reasonably well the experimental trends \cite{Speich1972} except perhaps for alloys encompassing about $10\%$ solute. This discrepancy is most likely due to the fact that the Fe-Mn and Fe-Ni binary alloys with $x\gtrsim 0.05$ exist as a mixture of bcc and fcc phases. On the theory side, the lattice parameters of Fe-Ni and Fe-Mn alloys are underestimated as a result of the employed exchange-correlation approximation (Fig. \ref{fig2}), which at least partly explains why theory in general overestimates the elastic moduli.

On a qualitative level, the trends of the bulk moduli we understand as the result of the interplay between chemical and volume effects. The calculated bulk moduli of pure Ni and Mn in the bcc lattice ($B$(Mn)$\approx 222$ GPa, $B$(Ni)$\approx 193$ GPa) are both larger than that of Fe. Hence, at large concentrations the bulk moduli of binary alloys should eventually increase. On the other hand, at low concentrations (less than $\sim 7\%$), the interaction between the impurity atoms is weak and thus the trend of the bulk modulus is mainly governed by the volume effect. Increasing volume in turn produces a drop in the bulk modulus, in agreement with Fig.~\ref{fig5}.

The results for Fe-Ni indicate that the polycrystalline elastic moduli $E$ and $G$ monotonically decrease with solute concentration. For Fe-Mn, the alloying effect can be divided into two parts: when $x$ is less than $\sim 0.05$, $G$ and $E$ remain constant with $x$. When $x$ is larger than $\sim 0.05$, both $G$ and $E$ increase with Mn content. Since the Voigt and Reuss bounds depend only on the single-crystal shear elastic constants, the trends of $G$ in Fig. \ref{fig5} directly emerge from those of $C_{44}$ and $C^\prime$ (Fig. \ref{fig4}). We recall that for isotropic crystals, $G$ reduces exactly to the single-crystal shear elastic constant ($C_{44}=C^\prime$). The Young modulus is a mixture of $B$ and the Pugh ratio $B/G$, \emph{viz.} $E=9B/(3B/G+1)$, which is nicely reflected by the trends in Fig. \ref{fig5}. For Fe$_{1-x}$Ni$_x$ with Ni content below $\sim 6\%$, $B/G$ remains nearly constant with $x$ (Table  \ref{tabII}), and thus the corresponding $E$ resembles the bulk modulus. At larger Ni concentrations, $B/G$ increases substantially which explain the continuous decrease of $E$. In the case of Fe-Mn, $B/G$ decreases with Mn addition which gives a strong positive slope to the Young modulus as compared to that of $B$.

\begin{figure}[b]
\includegraphics[scale=0.3]{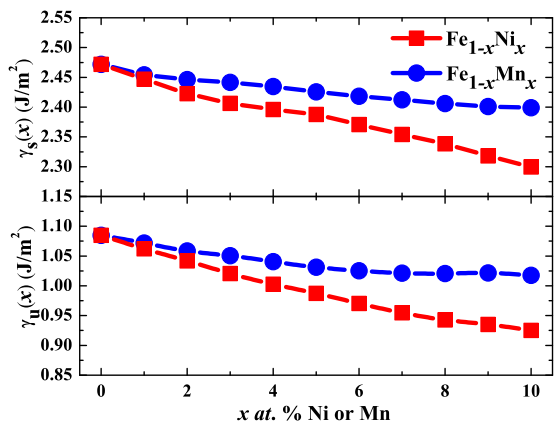}
\caption{(Color online) Surface energy and unstable stacking fault energy of bcc Fe$_{1-x}$Mn$_x$ and Fe$_{1-x}$Ni$_x$ alloys as a function of composition. The surface energy corresponds to the $\{110\}$ facet and the USF energy to the $\{110\}\langle111\rangle$ slip system.}\label{fig6}
\end{figure}

\begin{table*}
\caption{Theoretical single-crystal elastic constants $C_{11}(x)$, $C_{12}(x)$, $C_{44}(x)$, $C^\prime(x)$, polycrystalline bulk $B(x)$, shear $G(x)$ and Young $E(x)$ moduli (GPa), Poisson ratio $\nu$, Pugh ductile/brittle ratio $B/G$, Cauchy pressure $(C_{12}-C_{44})$ (GPa) and  Zener anisotropy ratio $C_{44}/C^\prime$ for Fe$_{1-x}$Mn$_x$ (upper panel) and Fe$_{1-x}$Ni$_x$ (lower panel) alloys as a function of composition.}\label{tabII}
\begin{ruledtabular}
\begin{tabular}{lcccccccccccc}
$x$ & $B$ & $C^\prime$ & $C_{11}$ & $C_{12}$ & $C_{44}$ & $G$ & $E$ &   $\nu$ & $(C_{12}-C_{44})$& $B/G$  & $C_{44}/C^\prime$\\
\hline
\multicolumn{12}{c}{Fe$_{1-x}$Mn$_x$}\\
\hline
   0&  185.4 & 77.6 & 288.9 & 133.7 & 100.0 & 90.3 &  233.1 & 0.290 & 33.71 & 2.05 & 1.29\\
0.01&  179.1 & 76.0 & 280.5 & 128.4 & 100.5 & 89.9 &  231.0 & 0.285 & 27.94 & 1.99 & 1.32\\
0.02&  172.0 & 75.2 & 272.3 & 121.8 & 102.8 & 90.7 &  231.4 & 0.276 & 19.06 & 1.90 & 1.37\\
0.03&  166.8 & 74.1 & 265.6 & 117.4 & 105.4 & 91.5 &  232.1 & 0.268 & 12.01 & 1.82 & 1.42\\
0.04&  161.0 & 73.3 & 258.8 & 112.1 & 107.7 & 92.3 &  232.5 & 0.259 &  4.41 & 1.74 & 1.47\\
0.05&  159.7 & 73.1 & 257.2 & 111.0 & 110.4 & 93.6 &  234.8 & 0.255 &  0.59 & 1.71 & 1.51\\
0.06&  155.4 & 73.6 & 253.5 & 106.3 & 113.0 & 95.2 &  237.1 & 0.246 & -6.71 & 1.63 & 1.54\\
0.07&  153.1 & 74.6 & 252.5 & 103.4 & 115.1 & 96.7 &  239.6 & 0.239 &-11.68 & 1.58 & 1.54\\
0.08&  153.6 & 75.6 & 254.4 & 103.2 & 116.8 & 98.1 &  242.6 & 0.237 &-13.55 & 1.57 & 1.54\\
0.09&  157.7 & 76.3 & 259.4 & 106.8 & 117.9 & 99.0 &  245.7 & 0.240 &-11.07 & 1.59 & 1.55\\
 0.1&  163.5 & 76.8 & 265.9 & 112.3 & 119.0 & 99.8 &  248.9 & 0.246 & -6.71 & 1.64 & 1.55\\
\hline
\multicolumn{12}{c}{Fe$_{1-x}$Ni$_x$}\\
\hline
   0&  185.4 & 77.6 & 288.9 & 133.7 & 100.0 & 90.3 & 233.1 & 0.290 & 33.7 & 2.05 & 1.29\\
0.01&  183.6 & 74.6 & 283.0 & 133.9 &  97.3 & 87.5 & 226.5 & 0.294 & 36.6 & 2.10 & 1.31\\
0.02&  181.2 & 72.6 & 278.0 & 132.8 &  96.6 & 86.1 & 223.1 & 0.295 & 36.2 & 2.10 & 1.33\\
0.03&  177.7 & 70.6 & 271.8 & 130.6 &  96.3 & 85.0 & 220.0 & 0.294 & 34.4 & 2.09 & 1.36\\
0.04&  174.6 & 68.7 & 266.1 & 128.8 &  96.4 & 84.2 & 217.5 & 0.292 & 32.4 & 2.07 & 1.40\\
0.05&  173.3 & 66.8 & 262.4 & 128.8 &  97.0 & 83.5 & 215.9 & 0.292 & 31.7 & 2.07 & 1.45\\
0.06&  173.6 & 64.9 & 260.2 & 130.3 &  98.2 & 83.2 & 215.2 & 0.293 & 32.1 & 2.09 & 1.51\\
0.07&  176.8 & 62.9 & 260.6 & 134.9 &  99.4 & 82.7 & 214.6 & 0.298 & 35.5 & 2.14 & 1.58\\
0.08&  181.8 & 60.7 & 262.8 & 141.3 & 100.6 & 82.2 & 214.2 & 0.304 & 40.7 & 2.21 & 1.66\\
0.09&  188.3 & 58.6 & 266.4 & 149.3 & 102.2 & 81.7 & 214.2 & 0.310 & 47.1 & 2.30 & 1.74\\
 0.1&  196.9 & 56.4 & 272.1 & 159.3 & 103.4 & 81.1 & 213.9 & 0.319 & 55.9 & 2.43 & 1.83\\
\end{tabular}
\end{ruledtabular}
\end{table*}

\subsection{Surface energy and unstable stacking fault energy}

The calculated surface energies and USF energies of Fe$_{1-x}$Mn$_x$ and Fe$_{1-x}$Ni$_x$ are shown in Fig.~\ref{fig6} as a function of $x$. Both Mn and Ni are predicted to decrease the surface energy of the $\{110\}$ facet of bcc Fe. Nickel has a stronger effect than Mn. Namely, $10 \%$ Ni addition reduces the surface energy of Fe by $0.17$\,J/m$^2$, which is about $7 \%$ of the surface energy of pure Fe, whereas Mn lowers the surface energy by $0.07$ J/m$^2$ (about $3 \%$). The alloying effect on the surface energy can be understood on a qualitative level using the surface energies calculated for the alloy components \cite{Punkkinen2011}. Nickel has substantially smaller surface energy (considering the close-packed fcc surfaces) than bcc Fe, and thus Ni addition is expected to reduce $\gamma_s$ of Fe. The surface energy of $\alpha-$Mn is larger than that of Fe \cite{Punkkinen2011a}. However, when considering the bcc lattice, the surface energy of Mn turns out to be intermediate between those of Ni and Fe, which is nicely reflected by the relative effects of Mn and Ni on $\gamma_s$.

The USF energy shows a similar dependence on Ni/Mn alloying as the surface energy; it decreases from $1.08$ to $0.92$ J/m$^2$ upon $10\%$ Ni addition, which is about 15\,\% of the USF energy of pure Fe. Compared to the effect of Ni, $10\%$ Mn produces a smaller change in $\gamma_u$ ($5.6 \%$). Using the same methodology as for Fe and Fe-alloy, we computed the USF of hypothetical bcc Mn and bcc Ni for the present slip system and at the volume of bcc Fe. We obtained $0.63$\,J/m$^2$ for Mn and $0.44$\,J/m$^2$ for Ni. These figures are in line with the trends from Fig. \ref{fig6}.

\begin{figure}
\includegraphics[scale=0.3]{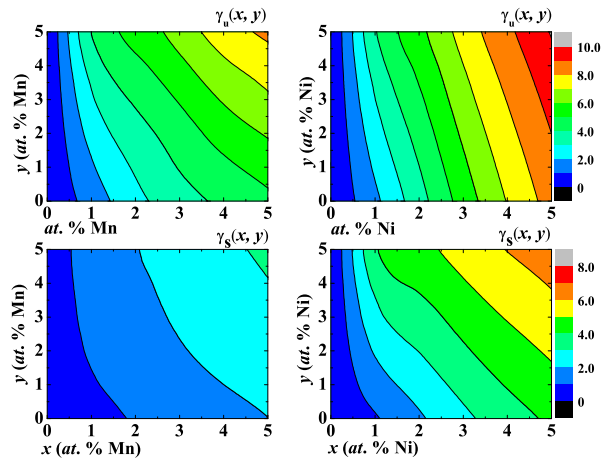}
\caption{(Color online) Relative change ($-[\gamma(x,y)-\gamma(0,0)]/\gamma(0,0)$) of the USF energy and the surface energy as a function of bulk concentration ($x$) of Mn/Ni (abscissae) and the additional solute concentration ($y$) at the fault plane (ordinates) for Fe$_{1-x}$Mn$_x$ and Fe$_{1-x}$Ni$_x$ ($ 0\leq x \leq 5\% $) alloys.}\label{fig7}
\end{figure}

The segregation effect on the surface energy and USF energy are shown in Fig.~\ref{fig7}. We find that both Mn and Ni segregate to the fault plane. Above $\sim 1\%$ solute concentration in bulk Fe, the surface energy and the USF energy decrease slightly with segregation of Ni/Mn to the layer at the planar fault. In this segregation calculation we kept the volume the same as in the bulk (host alloy). Therefore, only the chemical segregation effect is considered and the local volume expansion effect nearby fault plane due to the segregating solute atoms was ignored. The results show that the segregation effect on the surface energy is larger for Ni than for Mn. Furthermore, the surface segregation effect slightly increases with increasing Mn content in Fe-Mn dilute alloy. For the USF energy, the segregation effect of Mn is similar to that of Ni when the concentrations are small. However, with increasing solution amount $x$, the segregation of Mn to stacking fault increases with high degree, but the effect of Ni remains almost the same.

The surface energy and the USF energy are primarily determined by the properties of the surface layer and the slip layers, respectively. Segregation changes the concentration of the solute at the surface or slip plane. Because Mn and Ni have lower planar fault energies than Fe (in the bcc structure), both formation energies of the chemically homogeneous Fe-alloys are expected to decrease as the Mn or Ni solute concentration increases, in line with the present results. The effect is more pronounced for Ni since its USF/surface energy is smaller than that of Mn.

\section{Discussion}\label{discussion}

\subsection{Ductile and brittle properties}

The fracture behavior of a specific material rests on various conditions, such as, the presence of flaws, the way and the magnitude of the applied stress, temperature, strain rate, and alloying elements. Alloying changes the interaction among atoms, and produces indispensable influences on the mechanical properties like the elastic response or the ductile versus brittle behavior. In the following, the ductility of the dilute Fe-Mn and Fe-Ni binary alloys is addressed using previously established effective theoretical models and phenomenological relationships based on planar fault energies and elastic constants. These models are widely used and often overused in combination with \emph{ab initio} calculations and thus a careful assessment of their performance is of common interest. Here we employ four criteria discussed in the introduction to estimate the effect of Mn/Ni on the ductility of ferromagnetic bcc Fe. These criteria are based on the Poisson ratio ($\nu$), the Cauchy pressure ($C_{12}-C_{44}$), the Pugh ratio $B/G$, and the Rice ratio $\gamma_s/\gamma_u$. We recall that the first two criteria are closely connected with the Pugh conditions and thus no substantial deviations between them are expected.

\begin{figure}
\includegraphics[scale=0.3]{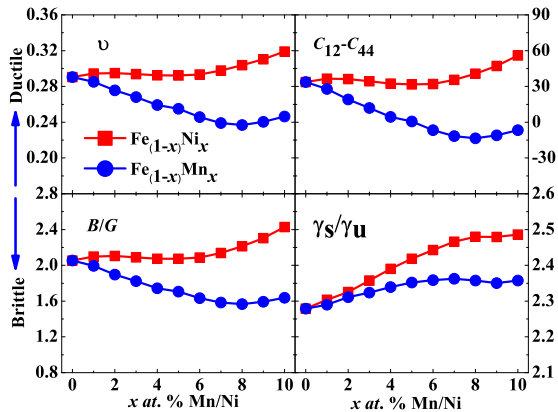}
\caption{(Color online) Poisson ratio $\nu$, Cauchy pressure $(C_{12}-C_{44})$ (in units of GPa), Pugh ratio $B/G$, and Rice ratio $\gamma_s/\gamma_u$ of bcc Fe$_{1-x}$Mn$_x$ and Fe$_{1-x}$Ni$_x$ alloys as a function of composition.}\label{fig8}
\end{figure}

The present theoretical predictions are shown in Fig.~\ref{fig8}. Both $B/G$ and Poisson's ratio indicate that Mn makes the ferrite Fe-based alloys more brittle. According to Pugh criterion, about $4 \%$ Mn is needed to transfer the Fe alloy from the ductile into the brittle regime. On the other hand, small Ni addition ($\lesssim 6 \%$) keeps the good ductility of Fe  whereas larger amounts of Ni make the Fe-Ni system more ductile.

The Cauchy pressure $(C_{12}-C_{44})$ follows very closely the trend of $B/G$ and that of the Poisson ratio. However, the Cauchy pressure becomes negative at a slightly larger concentration ($\sim 6\%$) than the critical value in terms of $B/G$ ($\sim 4\%$). The small difference between the "predictions" based on these phenomenological correlations originates from the elastic anisotropy of the present alloys (cf. Eq. \eqref{eq0}).

According to the theory developed by Rice~\cite{Rice1992}, the ratio $\gamma_s/\gamma_u$ is associated with the fracture behavior. The increment of $\gamma_s/\gamma_u$ stimulates the creation of dislocations. In such cases, the stress around a crack tip will be released by slipping the atomic layers. A decreasing ratio, on the other hand, means that the material will crack by opening new micro surfaces.

Figure~\ref{fig8} shows that for Fe$_{1-x}$Ni$_x$, $\gamma_s/\gamma_u$ increases as a function of $x$ indicating that the system becomes more ductile with Ni addition. Although Ni decreases both $\gamma_s$ and $\gamma_u$ simultaneously, the alloying effect on the USF energy is more pronounced, resulting in an increasing Rice ratio from 2.27 corresponding to pure Fe to 2.47 belonging to Fe$_{0.9}$Ni$_{0.1}$. In the case of Mn addition, we observe relatively small changes in the Rice ratio. Small amounts of Mn make the Fe-Mn alloy slightly more ductile in terms of the Rice parameter (compared to pure Fe), but with increasing Mn content beyond $\sim 6\%$ the $\gamma_s/\gamma_u$ ratio saturates around $2.35$.

According to the present theoretical results for the Rice parameter, Ni has stronger effect in making the alloy ductile compared to Mn. In addition, more than 6\% addition of Mn makes the bcc phase relatively more brittle. In fact, in terms of the USF, one would predict that both Ni and Mn make the dislocation formation in Fe more easy, and the effect of Ni is superior to that of Mn. On the other hand, both element decrease the surface energy as well, making the crack opening more likely.

Before turning to the comparison between the Pugh and Rice criteria, we make an observation based on the present segregation studies. Allowing for surface and interface segregation leads to small changes in the USF energy and the surface energy (cf. Section III.D). In terms of the Rice parameter, in alloys containing $5\%$ impurity $5\%$ surface/USF segregation increases slightly the $\gamma_s/\gamma_u$ ratio ($\sim 2\%$ for Fe-Ni and $\sim 4\%$ for Fe-Mn). That is because in both binary systems the USF energy is lowered by a larger degree than the surface energy upon segregation (Fig. \ref{fig7}). These changes are far below those associated with the effect of bulk concentration (Fig. \ref{fig8}). In addition, taking into account the different time-scales for atomic diffusion and dislocation movement, we conclude that the segregation effects may safely be omitted for the present discussion.

\subsection{Pugh criterion versus Rice criterion}

In the case of Fe-Ni alloys, the conclusion drawn from $\gamma_s/\gamma_u$ is consistent with the other three criteria based on $B/G$, Poisson ratio, and Cauchy pressure. When we look for the individual trends, we find that this consistency is to some extent incidental, especially at large Ni content. Namely, while both $G$ and $\gamma_u$ decrease with Ni addition, the trends for the surface energy and $B$ strongly deviate from each other above $\sim 5 \%$ Ni. At low Ni levels $B$ and $\gamma_s$ follow similar trends, but the relatively large $B$ of Fe$_{0.9}$Ni$_{0.1}$ is not supported by the strongly decreased $\gamma_s$ calculated for this alloy. The reason why $\gamma_s/\gamma_u$ and $B/G$ still predict similar effects (from the Rice and Pugh conditions) is simply due to the strong decrease calculated for $\gamma_u$ upon Ni addition.

In the case of Fe$_{1-x}$Mn$_x$, the conclusions made based on the Rice and Pugh criteria contradict each other. While $B/G$ decreases for $x\lesssim 0.08$, $\gamma_s/\gamma_u$ shows a weak increase for these alloys. We find that for this system, $G$ and $\gamma_u$ follow completely different trends as a function of Mn content (Figs. \ref{fig5} and \ref{fig6}), and the deviation between the trends of $\gamma_s$ and $B$ is also pronounced (although at low and intermediate Mn level both of them show decreasing trends). From these results, we conclude that the two ductility criteria (Pugh and Rice) lead to inconsistent results, a fact that questions their reliability and limits their scope.

In the following we make an attempt to understand the origin of this discrepancy. To this end, we make use of the particular shear elastic constant associated with the present slip system as well as of the concept of theoretical cleavage stress. These two quantities are considered here as possible alternative measures of the materials resistance to dislocation slip and cleavage, respectively, and are expected to give a better estimate of the corresponding effects compared to the polycrystalline $G$ and $B$ employed in the Pugh criterion.

\begin{figure}
\includegraphics[scale=0.32]{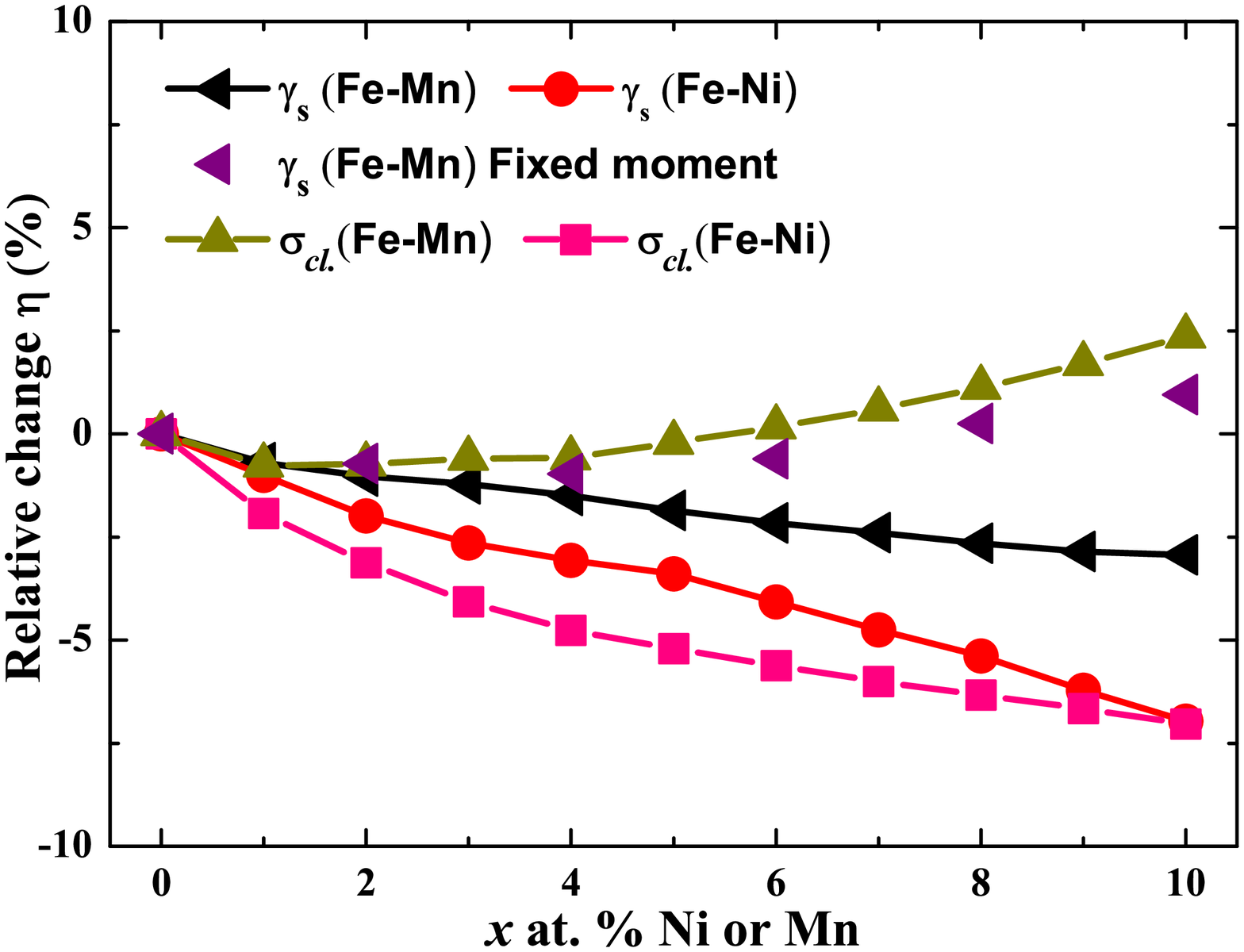}
\includegraphics[scale=0.32]{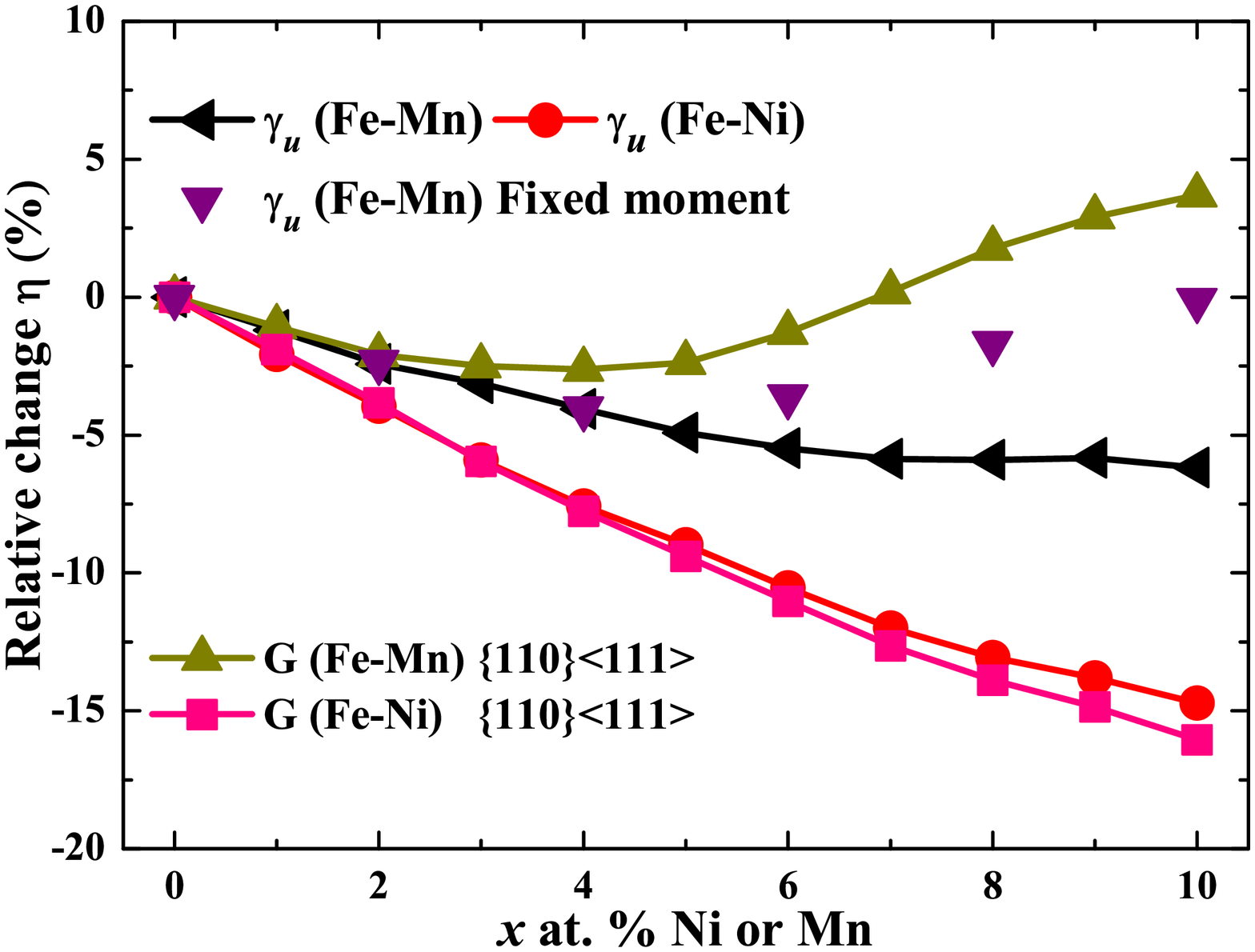}
\caption{(Color online) Relative changes ($\eta$ in $\%$) for the surface energy and the theoretical cleavage stress for the $\{110\}$ surface (upper panel) and for the USF energy and the single-crystal shear elastic modulus associated with the $1/2\langle111\rangle$ slip system (lower panel) for bcc Fe$_{1-x}$Mn$_x$ and Fe$_{1-x}$Ni$_x$ alloys as a function of composition. Fixed-spin results are shown for the surface energy and the USF energy of the Fe-Mn system.}\label{fig9}
\end{figure}

Within the Griffith theory of brittle fracture, the theoretical cleavage stress is often approximated as

\begin{eqnarray}
\sigma_{cl.}\{lmn\}=\left(\dfrac{E_{lmn}\gamma_{lmn}}{d_{lmn}}\right)^{1/2},
\end{eqnarray}
where $\{lmn\}$ stands for the cleavage plane, $E_{lmn}$, $\gamma_{lmn}$ and $d_{lmn}$ are the corresponding Young modulus, surface energy and interlayer distance, respectively. Irwing and Orowan extended the above equation (modified Griffith equation) by including the plastic work before fracture. Here we neglect this additional term, \emph{i.e.} $\gamma_{lmn}$ represents merely the solid vacuum interface energy. Using our calculated elastic constants, lattice parameters and surface energy, we computed $\sigma_{cl.}$ for Fe-Mn and Fe-Ni for the $\{110\}$ plane, for which $E_{110}=12C^{\prime}C_{44}B/(C_{11}C_{44}+3C^{\prime}B)$. The alloying-induced changes for $\sigma_{cl.}\{110\}$ are compared to those calculated for the surface energy in Fig.~\ref{fig9}, upper panel. Here the change $\eta_X(x) = [X(x)-X(0)]/X(0)$ ($X(x)$ stands for the cleavage stress or the surface energy for Fe$_{1-x}$M$_x$ alloy) is expressed relative to the corresponding value in pure Fe $X(0)$. It is found that $\eta_{\sigma_{cl.}}(x)$ and $\eta_{\gamma_s}$ follow similar trends for Fe-Ni, but strongly deviate for Fe-Mn. Before explaining this deviation in the case of Mn doping, we introduce the shear modulus associated with the present slip system.

\begin{figure}
	\includegraphics[scale=0.32]{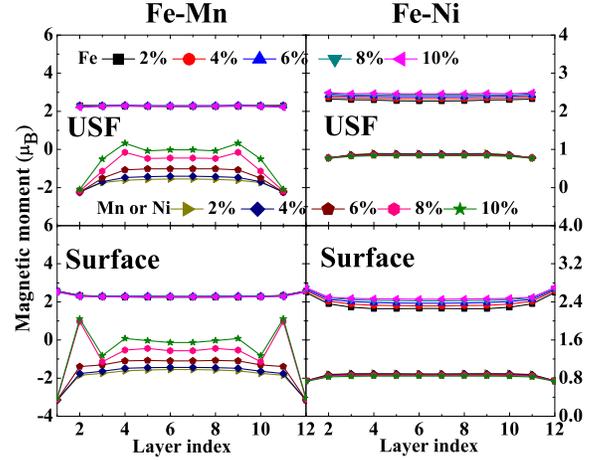}
	\caption{(Color online) Local magnetic moments of Fe (upper curves) and Mn/Ni (lower curves) for Fe$_{1-x}$Mn$_x$ (left panels) and Fe$_{1-x}$Ni$_x$ (right panels) as a function of layer index. Results are shown for the model systems used for the USF energy (upper panels) and surface energy (lower panels) calculations. Different symbols correspond to various impurity levels ($x$) as shown in the legend.}\label{fig10}
\end{figure}

In bcc alloys, slip occurs primarily in the $\{110\}$ plane along the $\langle111\rangle$ with Burgers vector $(1/2,1/2,1/2)$. The associated shear modulus can be expressed as

\begin{eqnarray}
G\{110\}\langle111\rangle=\dfrac{3C_{44}C^\prime}{C^\prime+2C_{44}}.
\end{eqnarray}
We note that the above modulus in fact expresses the shear for any possible shear plane $\{lmn\}$ which contains the $\langle111\rangle$ shear direction. In the original Pugh criteria, the averaged shear modulus $G$ is used which in anisotropic materials can substantially differ from $G\{110\}\langle111\rangle$. The alloying-induced changes for $G\{110\}\langle111\rangle$ are compared to those calculated for the USF energy in Fig.~\ref{fig9}, lower panel.
Interestingly, we find that $\eta_{G\{110\}\langle111\rangle}$ and $\eta_{\gamma_u}$ are practically identical for Fe-Ni. Both $G\{110\}\langle111\rangle$ and $\gamma_u$ decrease by $\sim 15\%$ when $10\%$ Ni is added to Fe. Nickel substantially softens the elastic modulus associated with the $\{110\}\langle111\rangle$ shear which is nicely reflected by the decrease of the energy barrier for the $\{110\}\langle111\rangle$ slip. For Fe-Mn, $\eta_{G\{110\}\langle111\rangle}$ and $\eta_{\gamma_u}$ are also close to each other for Mn content below $\sim 5\%$, but show large deviations at larger concentrations. Adding more Mn to Fe$_{0.95}$Mn$_{0.05}$ further decreases the USF energy, but $G\{110\}\langle111\rangle$ changes slope and shows a weak increase for Fe$_{0.9}$Mn$_{0.1}$ relative to the value for pure Fe.

We suggest that the different behaviors obtained for Fe-Ni and Fe-Mn has to a large extent magnetic origin. That is because in contrast to Ni, Mn is a weakly itinerant magnet and thus any (here structural) perturbation can have a marked impact on its magnetic state. To illustrate that, in Fig.~\ref{fig10} we plotted the local magnetic moments for the unit cells used for the USF energy and surface energy calculations. In the case of Fe-Ni, we see no substantial deviation in the local magnetic moments as we approach the planar fault area. Both Ni and Fe moments near the fault plane remain close to the bulk value. However, for Fe-Mn, the local magnetic moments of Mn next to the planar fault prefer a very strong antiferromagnetic coupling with the Fe matrix irrespective of the bulk concentration. Although the bulk moments approach zero as the Mn content increases to $10 \%$, the interface Mn moments remain around $-2.2 \mu_{\rm B}$ and those on the surface around $-3.1 \mu_{\rm B}$. This strong antiferromagnetic Fe-Mn coupling near the USF interface and surface indicates an energetically stable configuration that can lower the corresponding formation energy. Indeed, removing this degree of freedom by fixing all Mn moments to the corresponding bulk value (\emph{i.e.}, modeling a situation similar to the case of Fe-Ni system) increases the the surface energy and USF energy of Fe-Mn. The relative changes of the corresponding $\gamma_s^{\rm FS}$ and $\gamma_u^{\rm FS}$ values (FS stands for fixed-spin) relative to those of pure Fe are shown in Fig.~\ref{fig9}. It is found that $\gamma_u^{\rm FS}$ for Fe$_{0.9}$Mn$_{0.1}$ approaches the USF energy of pure Fe. In fact, constraining the magnetic moments near the planar fault brings the trend followed by $\gamma_u^{\rm FS}$ rather close to that of $G\{110\}\langle111\rangle$ (Fig. \ref{fig9}, lower panel). A similar impact of the constrained magnetic moment is seen for the surface energy as well. Namely, for Fe-Mn $\gamma_s^{\rm FS}$ and the theoretical cleavage stress show similar concentration dependencies (Fig.~\ref{fig9}, upper panel). On this ground, we conclude that the deviations seen between the surface energy and cleavage stress and between the USF energy and the shear modulus in the case of Fe-Mn are due to the stable antiferromagnetic state of Mn near the planar faults.

\begin{figure}
	\includegraphics[scale=0.32]{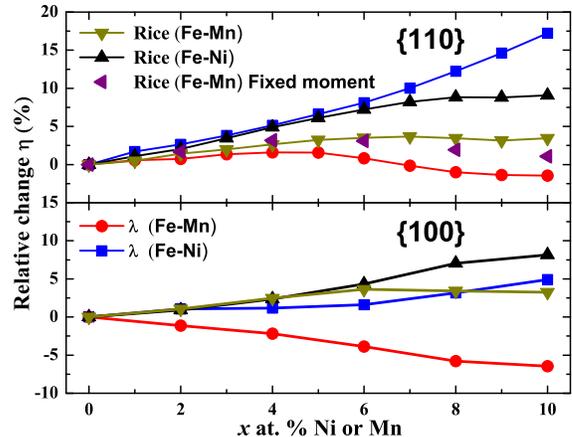}
	\caption{(Color online) Theoretical Rice parameter and the ratio between the theoretical cleavage stress and the shear modulus associated with the slip system ($\lambda$) for bcc Fe$_{1-x}$Mn$_x$ and Fe$_{1-x}$Ni$_x$ alloys as a function of composition. Fixed-spin results are shown for Rice parameter of the Fe-Mn system. Upper panel represents $\{110\}$ as the cleavage plane, and lower panel $\{100\}$ as the cleavage plane.}\label{fig11}
\end{figure}

Using the FS results for the Rice parameter, we get a weak decrease of $\gamma_s^{\rm FS}/\gamma_u^{\rm FS}$ above $\sim 5\%$ Mn addition to Fe (shown in Fig.~\ref{fig11}, upper panel). Thus, the fixed-moment result for the Rice parameter is somewhat closer to the original Pugh ratio, both of them predicting enhanced intrinsic brittleness for the Fe-Mn solid solution. This demonstrates that the discrepancy seen between the two criteria (Fig.~\ref{fig8}) is partly due to the weak magnetic behavior of Mn associated with the planar defects but absent in the elastic moduli. We suspect that the situation in non-magnetic alloys could be closer to the case of Fe-Ni, where a good parallelism between the Pugh and Rice conditions is found. However, further theoretical research is needed before making the final verdict for this question.

Finally, we test the ratio between the cleavage energy and the shear modulus associated with the slip system, $\lambda\{lmn\} \equiv \sigma_{cl.}\{lmn\}/G\{110\}\langle111\rangle$, as a possible alternative measure of the ductile-brittle behavior. In the upper panel of Fig.~\ref{fig11}, we compare the Rice parameters (from Fig.~\ref{fig9}) to $\lambda\{110\}$. For Fe-Ni, we find a good correspondence between the two measures. Namely both of them increase with Ni addition, indicating enhanced ductility. For Fe-Mn, slightly larger deviation occurs at large Mn content. As we discussed above, part of this deviation may be ascribed to magnetism and in particular to the stable antiferromagnetic state of Mn around the planar faults.

Since in bcc metals, cleavage predominantly occurs in the $\{100\}$ plane, in addition to the previously discussed cleavage stress, we also computed $\sigma_{cl.}\{100\}$ using the corresponding surface energy and
$E_{100}=6 C^{\prime}B/(C_{11}+C_{12})$. The so obtained $\lambda\{100\}$ parameters are shown in the lower panel of Fig.~\ref{fig11} together with the Rice parameters calculated using the surface energy for the $\{100\}$ facet. We find that the two sets of Rice parameters (upper and lower panels) predict rather similar effects. The situation for $\lambda$ is very different: $\lambda\{100\}$ monotonously increases with Ni and decreases with Mn addition. The inconsistency between the Rice parameter and $\lambda$ obtained for the $\{100\}$ plane is a consequence of the limited information built in the Rice ratio (merely through the surface energy) compared to the cleavage stress (involving both the Young modulus and the surface energy). We conclude that in contrast to the Rice parameter, the average $\lambda$ (taking into account both cleavage planes) indicates increased brittleness for Fe-Mn and increased ductility for Fe-Ni. For both alloys, the present predictions based on $\lambda$ are in line with the observations~\cite{Wessman2008}.

\section{Conclusion}\label{conclusion}

We have investigated the effect of Mn and Ni addition on the mechanical properties of ferromagnetic bcc Fe. Fe-Ni, the lattice parameter increases nearly linearly with alloying whereas in Fe-Mn, the lattice parameter increases up to $\sim5\,\%$ Mn and then decreases with further Mn addition.

For both Fe-Mn and Fe-Ni alloys, the elastic moduli show a nonlinear trends as a function of concentration. The surface energy and the unstable stacking fault energy decrease by adding Mn or Ni to Fe. For both planar fault energies, Ni shows a stronger effect than Mn. Segregation seems to have a minor effect on the surface and USF energies for dilute Fe-Ni and Fe-Mn alloys. According to semi-empirical correlations based on the Poisson ratio, the Cauchy pressure, and the Pugh ratio, we have found that Mn makes the bcc Fe-based dilute alloy more brittle and Ni makes it more ductile. However, the study of $\gamma_s/\gamma_u$ for the $\{110\}$ surface and the $\{110\}\langle111\rangle$ slip systems indicates that both binary alloys become more ductile with increasing solute concentration although the relative fracture behavior is consistent with the other three criteria. This discrepancy between the Rice and the Pugh criteria  is ascribed to the complex magnetic effects around the planar fault, which are however missing from the bulk parameters entering the Pugh/Cauchy criterion. 

Using the theoretical cleavage stress and the shear modulus associated with the dominant slip system in bcc alloys, we introduce an alternative measure for the ductile-brittle behavior. We find that our $\lambda\{100\}\equiv \sigma_{cl.}\{100\}/G\{110\}\langle111\rangle$ ratio is able to capture the observed alloying effects in the mechanical properties of Fe-rich Fe-Ni and Fe-Mn alloys. The superior performance of $\lambda$ compared to the Rice parameter lies in the additional information built in the theoretical cleavage stress as compared the surface energy.

\emph{Acknowledgements}
Work supported by the Swedish Research Council, the Swedish Foundation for Strategic Research, and the China Scholarship Council. The National 973 Project of China (Grant No. 2014CB644001) and the Hungarian Scientific Research Fund (OTKA 84078 and 109570) are also acknowledged for financial support.


\begin{thebibliography}{62}
	\expandafter\ifx\csname natexlab\endcsname\relax\def\natexlab#1{#1}\fi
	\expandafter\ifx\csname bibnamefont\endcsname\relax
	\def\bibnamefont#1{#1}\fi
	\expandafter\ifx\csname bibfnamefont\endcsname\relax
	\def\bibfnamefont#1{#1}\fi
	\expandafter\ifx\csname citenamefont\endcsname\relax
	\def\citenamefont#1{#1}\fi
	\expandafter\ifx\csname url\endcsname\relax
	\def\url#1{\texttt{#1}}\fi
	\expandafter\ifx\csname urlprefix\endcsname\relax\def\urlprefix{URL }\fi
	\providecommand{\bibinfo}[2]{#2}
	\providecommand{\eprint}[2][]{\url{#2}}
	
	\bibitem[{\citenamefont{Wessman et~al.}(2008)\citenamefont{Wessman, Hertzman,
			Pettersson, Lagneborg, and Liljas}}]{Wessman2008}
	\bibinfo{author}{\bibfnamefont{S.}~\bibnamefont{Wessman}},
	\bibinfo{author}{\bibfnamefont{S.}~\bibnamefont{Hertzman}},
	\bibinfo{author}{\bibfnamefont{R.}~\bibnamefont{Pettersson}},
	\bibinfo{author}{\bibfnamefont{R.}~\bibnamefont{Lagneborg}},
	\bibnamefont{and} \bibinfo{author}{\bibfnamefont{M.}~\bibnamefont{Liljas}},
	\bibinfo{journal}{Mater. Sci. Technol.} \textbf{\bibinfo{volume}{24}},
	\bibinfo{pages}{348} (\bibinfo{year}{2008}).
	
	\bibitem[{\citenamefont{Pugh}(1954)}]{Pugh1954}
	\bibinfo{author}{\bibfnamefont{S.}~\bibnamefont{Pugh}},
	\bibinfo{journal}{Philos. Mag.} \textbf{\bibinfo{volume}{45}},
	\bibinfo{pages}{823} (\bibinfo{year}{1954}).
	
	\bibitem[{\citenamefont{Rice}(1992)}]{Rice1992}
	\bibinfo{author}{\bibfnamefont{J.~R.} \bibnamefont{Rice}}, \bibinfo{journal}{J.
		Mech. Phys. Solids} \textbf{\bibinfo{volume}{40}}, \bibinfo{pages}{239}
	(\bibinfo{year}{1992}).
	
	\bibitem[{\citenamefont{Kumar et~al.}(2013)\citenamefont{Kumar, Tewari,
			Durgaprasad, Dutta, and Dey}}]{Kumar2013}
	\bibinfo{author}{\bibfnamefont{N.~N.} \bibnamefont{Kumar}},
	\bibinfo{author}{\bibfnamefont{R.}~\bibnamefont{Tewari}},
	\bibinfo{author}{\bibfnamefont{P.}~\bibnamefont{Durgaprasad}},
	\bibinfo{author}{\bibfnamefont{B.}~\bibnamefont{Dutta}}, \bibnamefont{and}
	\bibinfo{author}{\bibfnamefont{G.}~\bibnamefont{Dey}},
	\bibinfo{journal}{Comp. Mat. Sci.} \textbf{\bibinfo{volume}{77}},
	\bibinfo{pages}{260} (\bibinfo{year}{2013}).
	
	\bibitem[{\citenamefont{Vitos et~al.}(1994)\citenamefont{Vitos, Koll{\'a}r, and
			Skriver}}]{Vitos1994}
	\bibinfo{author}{\bibfnamefont{L.}~\bibnamefont{Vitos}},
	\bibinfo{author}{\bibfnamefont{J.}~\bibnamefont{Koll{\'a}r}},
	\bibnamefont{and} \bibinfo{author}{\bibfnamefont{H.~L.}
		\bibnamefont{Skriver}}, \bibinfo{journal}{Phys. Rev. B}
	\textbf{\bibinfo{volume}{49}}, \bibinfo{pages}{16694} (\bibinfo{year}{1994}).
	
	\bibitem[{\citenamefont{Vitos et~al.}(1997)\citenamefont{Vitos, Koll{\'a}r, and
			Skriver}}]{Vitos1997}
	\bibinfo{author}{\bibfnamefont{L.}~\bibnamefont{Vitos}},
	\bibinfo{author}{\bibfnamefont{J.}~\bibnamefont{Koll{\'a}r}},
	\bibnamefont{and} \bibinfo{author}{\bibfnamefont{H.~L.}
		\bibnamefont{Skriver}}, \bibinfo{journal}{Phys. Rev. B}
	\textbf{\bibinfo{volume}{55}}, \bibinfo{pages}{4947} (\bibinfo{year}{1997}).
	
	\bibitem[{\citenamefont{Koll{\'a}r et~al.}(2000)\citenamefont{Koll{\'a}r,
			Vitos, Skriver, and Dreyss{\'e}}}]{Kollar2000}
	\bibinfo{author}{\bibfnamefont{J.}~\bibnamefont{Koll{\'a}r}},
	\bibinfo{author}{\bibfnamefont{L.}~\bibnamefont{Vitos}},
	\bibinfo{author}{\bibfnamefont{H.}~\bibnamefont{Skriver}}, \bibnamefont{and}
	\bibinfo{author}{\bibfnamefont{H.}~\bibnamefont{Dreyss{\'e}}},
	\bibinfo{journal}{Lecture Notes in Physics (Springer-Verlag, Berlin, 2000)}
	(\bibinfo{year}{2000}).
	
	\bibitem[{\citenamefont{Kwon et~al.}(2005)\citenamefont{Kwon, Nabi, K{\'a}das,
			Vitos, Koll{\'a}r, Johansson, and Ahuja}}]{Kwon2005}
	\bibinfo{author}{\bibfnamefont{S.}~\bibnamefont{Kwon}},
	\bibinfo{author}{\bibfnamefont{Z.}~\bibnamefont{Nabi}},
	\bibinfo{author}{\bibfnamefont{K.}~\bibnamefont{K{\'a}das}},
	\bibinfo{author}{\bibfnamefont{L.}~\bibnamefont{Vitos}},
	\bibinfo{author}{\bibfnamefont{J.}~\bibnamefont{Koll{\'a}r}},
	\bibinfo{author}{\bibfnamefont{B.}~\bibnamefont{Johansson}},
	\bibnamefont{and} \bibinfo{author}{\bibfnamefont{R.}~\bibnamefont{Ahuja}},
	\bibinfo{journal}{Phys. Rev. B} \textbf{\bibinfo{volume}{72}},
	\bibinfo{pages}{235423} (\bibinfo{year}{2005}).
	
	\bibitem[{\citenamefont{Ald\'en et~al.}(1992)\citenamefont{Ald\'en, Skriver,
			Mirbt, and Johansson}}]{PRL69}
	\bibinfo{author}{\bibfnamefont{M.}~\bibnamefont{Ald\'en}},
	\bibinfo{author}{\bibfnamefont{H.~L.} \bibnamefont{Skriver}},
	\bibinfo{author}{\bibfnamefont{S.}~\bibnamefont{Mirbt}}, \bibnamefont{and}
	\bibinfo{author}{\bibfnamefont{B.}~\bibnamefont{Johansson}},
	\bibinfo{journal}{Phys. Rev. Lett.} \textbf{\bibinfo{volume}{69}},
	\bibinfo{pages}{2296} (\bibinfo{year}{1992}).
	
	\bibitem[{\citenamefont{Vitos}(2007)}]{Vitos-book}
	\bibinfo{author}{\bibfnamefont{L.}~\bibnamefont{Vitos}},
	\emph{\bibinfo{title}{Computational Quantum Mechanics for Materials
			Engineers: The EMTO Method and Applications}} (\bibinfo{publisher}{Springer},
	\bibinfo{year}{2007}).
	
	\bibitem[{\citenamefont{Pettifor}(1992)}]{Pettifor1992}
	\bibinfo{author}{\bibfnamefont{D.}~\bibnamefont{Pettifor}},
	\bibinfo{journal}{Mater. Sci. Technol.} \textbf{\bibinfo{volume}{8}},
	\bibinfo{pages}{345} (\bibinfo{year}{1992}).
	
	\bibitem[{\citenamefont{Vitos et~al.}(2000)\citenamefont{Vitos, Skriver,
			Johansson, and Koll{\'a}r}}]{Vitos2000}
	\bibinfo{author}{\bibfnamefont{L.}~\bibnamefont{Vitos}},
	\bibinfo{author}{\bibfnamefont{H.~L.} \bibnamefont{Skriver}},
	\bibinfo{author}{\bibfnamefont{B.}~\bibnamefont{Johansson}},
	\bibnamefont{and}
	\bibinfo{author}{\bibfnamefont{J.}~\bibnamefont{Koll{\'a}r}},
	\bibinfo{journal}{Comput. Mater. Sci.} \textbf{\bibinfo{volume}{18}},
	\bibinfo{pages}{24} (\bibinfo{year}{2000}).
	
	\bibitem[{\citenamefont{Vitos}(2001)}]{Vitos2001}
	\bibinfo{author}{\bibfnamefont{L.}~\bibnamefont{Vitos}},
	\bibinfo{journal}{Phys. Rev. B} \textbf{\bibinfo{volume}{64}},
	\bibinfo{pages}{014107} (\bibinfo{year}{2001}).
	
	\bibitem[{\citenamefont{Vitos et~al.}(2001)\citenamefont{Vitos, Abrikosov, and
			Johansson}}]{Vitos-PRL}
	\bibinfo{author}{\bibfnamefont{L.}~\bibnamefont{Vitos}},
	\bibinfo{author}{\bibfnamefont{I.}~\bibnamefont{Abrikosov}},
	\bibnamefont{and}
	\bibinfo{author}{\bibfnamefont{B.}~\bibnamefont{Johansson}},
	\bibinfo{journal}{Phys. Rev. Lett.} \textbf{\bibinfo{volume}{87}},
	\bibinfo{pages}{156401} (\bibinfo{year}{2001}).
	
	\bibitem[{\citenamefont{Andersen et~al.}(1998)\citenamefont{Andersen,
			Arcangeli, Tank, Saha-Dasgupta, Krier, Jepsen, and Dasgupta}}]{Andersen1998}
	\bibinfo{author}{\bibfnamefont{O.}~\bibnamefont{Andersen}},
	\bibinfo{author}{\bibfnamefont{C.}~\bibnamefont{Arcangeli}},
	\bibinfo{author}{\bibfnamefont{R.}~\bibnamefont{Tank}},
	\bibinfo{author}{\bibfnamefont{T.}~\bibnamefont{Saha-Dasgupta}},
	\bibinfo{author}{\bibfnamefont{G.}~\bibnamefont{Krier}},
	\bibinfo{author}{\bibfnamefont{O.}~\bibnamefont{Jepsen}}, \bibnamefont{and}
	\bibinfo{author}{\bibfnamefont{I.}~\bibnamefont{Dasgupta}}, in
	\emph{\bibinfo{booktitle}{Mat. Res. Soc. Symp. Proc}} (\bibinfo{year}{1998}),
	pp. \bibinfo{pages}{3--34}.
	
	\bibitem[{\citenamefont{Zwierzycki and Andersen}(2008)}]{Zwierzycki2009}
	\bibinfo{author}{\bibfnamefont{M.}~\bibnamefont{Zwierzycki}} \bibnamefont{and}
	\bibinfo{author}{\bibfnamefont{O.}~\bibnamefont{Andersen}},
	\bibinfo{journal}{Acta Phys. Pol., A}  (\bibinfo{year}{2008}).
	
	\bibitem[{\citenamefont{Soven}(1967)}]{Soven1967}
	\bibinfo{author}{\bibfnamefont{P.}~\bibnamefont{Soven}},
	\bibinfo{journal}{Phys. Rev.} \textbf{\bibinfo{volume}{156}},
	\bibinfo{pages}{809} (\bibinfo{year}{1967}).
	
	\bibitem[{\citenamefont{Gyorffy}(1972)}]{Gyorffy1972}
	\bibinfo{author}{\bibfnamefont{B.}~\bibnamefont{Gyorffy}},
	\bibinfo{journal}{Phys. Rev. B} \textbf{\bibinfo{volume}{5}},
	\bibinfo{pages}{2382} (\bibinfo{year}{1972}).
	
	\bibitem[{\citenamefont{Koll{\'a}r et~al.}(1997)\citenamefont{Koll{\'a}r,
			Vitos, and Skriver}}]{Kollar1997}
	\bibinfo{author}{\bibfnamefont{J.}~\bibnamefont{Koll{\'a}r}},
	\bibinfo{author}{\bibfnamefont{L.}~\bibnamefont{Vitos}}, \bibnamefont{and}
	\bibinfo{author}{\bibfnamefont{H.~L.} \bibnamefont{Skriver}},
	\bibinfo{journal}{Phys. Rev. B} \textbf{\bibinfo{volume}{55}},
	\bibinfo{pages}{15353} (\bibinfo{year}{1997}).
	
	\bibitem[{\citenamefont{Delczeg-Czirjak
			et~al.}(2011)\citenamefont{Delczeg-Czirjak, Nurmi, Kokko, and
			Vitos}}]{Delczeg2011}
	\bibinfo{author}{\bibfnamefont{E.}~\bibnamefont{Delczeg-Czirjak}},
	\bibinfo{author}{\bibfnamefont{E.}~\bibnamefont{Nurmi}},
	\bibinfo{author}{\bibfnamefont{K.}~\bibnamefont{Kokko}}, \bibnamefont{and}
	\bibinfo{author}{\bibfnamefont{L.}~\bibnamefont{Vitos}},
	\bibinfo{journal}{Phys. Rev. B} \textbf{\bibinfo{volume}{84}},
	\bibinfo{pages}{094205} (\bibinfo{year}{2011}).
	
	\bibitem[{\citenamefont{Li et~al.}(2014{\natexlab{a}})\citenamefont{Li, Li,
			Sch{\"o}necker, Li, Delczeg-Czirjak, Kvashnin, Eriksson, Johansson, and
			Vitos}}]{GJ}
	\bibinfo{author}{\bibfnamefont{G.}~\bibnamefont{Li}},
	\bibinfo{author}{\bibfnamefont{W.}~\bibnamefont{Li}},
	\bibinfo{author}{\bibfnamefont{S.}~\bibnamefont{Sch{\"o}necker}},
	\bibinfo{author}{\bibfnamefont{X.}~\bibnamefont{Li}},
	\bibinfo{author}{\bibfnamefont{E.~K.} \bibnamefont{Delczeg-Czirjak}},
	\bibinfo{author}{\bibfnamefont{Y.~O.} \bibnamefont{Kvashnin}},
	\bibinfo{author}{\bibfnamefont{O.}~\bibnamefont{Eriksson}},
	\bibinfo{author}{\bibfnamefont{B.}~\bibnamefont{Johansson}},
	\bibnamefont{and} \bibinfo{author}{\bibfnamefont{L.}~\bibnamefont{Vitos}},
	\bibinfo{journal}{Appl. Phys. Lett.} \textbf{\bibinfo{volume}{105}},
	\bibinfo{eid}{262405} (\bibinfo{year}{2014}{\natexlab{a}}).
	
	\bibitem[{\citenamefont{Landa et~al.}(2006)\citenamefont{Landa, Klepeis,
			S{\"o}derlind, Naumov, Velikokhatnyi, Vitos, and Ruban}}]{Landa2006}
	\bibinfo{author}{\bibfnamefont{A.}~\bibnamefont{Landa}},
	\bibinfo{author}{\bibfnamefont{J.}~\bibnamefont{Klepeis}},
	\bibinfo{author}{\bibfnamefont{P.}~\bibnamefont{S{\"o}derlind}},
	\bibinfo{author}{\bibfnamefont{I.}~\bibnamefont{Naumov}},
	\bibinfo{author}{\bibfnamefont{O.}~\bibnamefont{Velikokhatnyi}},
	\bibinfo{author}{\bibfnamefont{L.}~\bibnamefont{Vitos}}, \bibnamefont{and}
	\bibinfo{author}{\bibfnamefont{A.}~\bibnamefont{Ruban}}, \bibinfo{journal}{J.
		Phys.: Condens. Matter} \textbf{\bibinfo{volume}{18}}, \bibinfo{pages}{5079}
	(\bibinfo{year}{2006}).
	
	\bibitem[{\citenamefont{Magyari-K{\"o}pe
			et~al.}(2001)\citenamefont{Magyari-K{\"o}pe, Vitos, Johansson, and
			Koll{\'a}r}}]{Magyari2001}
	\bibinfo{author}{\bibfnamefont{B.}~\bibnamefont{Magyari-K{\"o}pe}},
	\bibinfo{author}{\bibfnamefont{L.}~\bibnamefont{Vitos}},
	\bibinfo{author}{\bibfnamefont{B.}~\bibnamefont{Johansson}},
	\bibnamefont{and}
	\bibinfo{author}{\bibfnamefont{J.}~\bibnamefont{Koll{\'a}r}},
	\bibinfo{journal}{Acta Crystallogr., Sect. B: Struct. Sci}
	\textbf{\bibinfo{volume}{57}}, \bibinfo{pages}{491} (\bibinfo{year}{2001}).
	
	\bibitem[{\citenamefont{Huang et~al.}(2006)\citenamefont{Huang, Vitos, Kwon,
			Johansson, and Ahuja}}]{Huang2006}
	\bibinfo{author}{\bibfnamefont{L.}~\bibnamefont{Huang}},
	\bibinfo{author}{\bibfnamefont{L.}~\bibnamefont{Vitos}},
	\bibinfo{author}{\bibfnamefont{S.}~\bibnamefont{Kwon}},
	\bibinfo{author}{\bibfnamefont{B.}~\bibnamefont{Johansson}},
	\bibnamefont{and} \bibinfo{author}{\bibfnamefont{R.}~\bibnamefont{Ahuja}},
	\bibinfo{journal}{Phys. Rev. B} \textbf{\bibinfo{volume}{73}},
	\bibinfo{pages}{104203} (\bibinfo{year}{2006}).
	
	\bibitem[{\citenamefont{Koll{\'a}r et~al.}(2003)\citenamefont{Koll{\'a}r,
			Vitos, Osorio-Guill{\'e}n, and Ahuja}}]{Kollar2003}
	\bibinfo{author}{\bibfnamefont{J.}~\bibnamefont{Koll{\'a}r}},
	\bibinfo{author}{\bibfnamefont{L.}~\bibnamefont{Vitos}},
	\bibinfo{author}{\bibfnamefont{J.}~\bibnamefont{Osorio-Guill{\'e}n}},
	\bibnamefont{and} \bibinfo{author}{\bibfnamefont{R.}~\bibnamefont{Ahuja}},
	\bibinfo{journal}{Phys. Rev. B} \textbf{\bibinfo{volume}{68}},
	\bibinfo{pages}{245417} (\bibinfo{year}{2003}).
	
	\bibitem[{\citenamefont{Magyari-K{\"o}pe
			et~al.}(2004)\citenamefont{Magyari-K{\"o}pe, Vitos, and
			Grimvall}}]{Magyari2004}
	\bibinfo{author}{\bibfnamefont{B.}~\bibnamefont{Magyari-K{\"o}pe}},
	\bibinfo{author}{\bibfnamefont{L.}~\bibnamefont{Vitos}}, \bibnamefont{and}
	\bibinfo{author}{\bibfnamefont{G.}~\bibnamefont{Grimvall}},
	\bibinfo{journal}{Phys. Rev. B} \textbf{\bibinfo{volume}{70}},
	\bibinfo{pages}{052102} (\bibinfo{year}{2004}).
	
	\bibitem[{\citenamefont{Hu et~al.}(2009)\citenamefont{Hu, Li, Yang, Kulkova,
			Bazhanov, Johansson, and Vitos}}]{Hu2009}
	\bibinfo{author}{\bibfnamefont{Q.-M.} \bibnamefont{Hu}},
	\bibinfo{author}{\bibfnamefont{C.-M.} \bibnamefont{Li}},
	\bibinfo{author}{\bibfnamefont{R.}~\bibnamefont{Yang}},
	\bibinfo{author}{\bibfnamefont{S.~E.} \bibnamefont{Kulkova}},
	\bibinfo{author}{\bibfnamefont{D.~I.} \bibnamefont{Bazhanov}},
	\bibinfo{author}{\bibfnamefont{B.}~\bibnamefont{Johansson}},
	\bibnamefont{and} \bibinfo{author}{\bibfnamefont{L.}~\bibnamefont{Vitos}},
	\bibinfo{journal}{Phys. Rev. B} \textbf{\bibinfo{volume}{79}},
	\bibinfo{pages}{144112} (\bibinfo{year}{2009}).
	
	\bibitem[{\citenamefont{Zhang et~al.}(2009)\citenamefont{Zhang, Johansson, and
			Vitos}}]{Zhang2009}
	\bibinfo{author}{\bibfnamefont{H.}~\bibnamefont{Zhang}},
	\bibinfo{author}{\bibfnamefont{B.}~\bibnamefont{Johansson}},
	\bibnamefont{and} \bibinfo{author}{\bibfnamefont{L.}~\bibnamefont{Vitos}},
	\bibinfo{journal}{Phys. Rev. B} \textbf{\bibinfo{volume}{79}},
	\bibinfo{pages}{224201} (\bibinfo{year}{2009}).
	
	\bibitem[{\citenamefont{Zander et~al.}(2007)\citenamefont{Zander,
			Sandstr{\"o}m, and Vitos}}]{Zander2007}
	\bibinfo{author}{\bibfnamefont{J.}~\bibnamefont{Zander}},
	\bibinfo{author}{\bibfnamefont{R.}~\bibnamefont{Sandstr{\"o}m}},
	\bibnamefont{and} \bibinfo{author}{\bibfnamefont{L.}~\bibnamefont{Vitos}},
	\bibinfo{journal}{Comput. Mater. Sci.} \textbf{\bibinfo{volume}{41}},
	\bibinfo{pages}{86} (\bibinfo{year}{2007}).
	
	\bibitem[{\citenamefont{Vitos et~al.}(2006)\citenamefont{Vitos, Korzhavyi, and
			Johansson}}]{Vitos2006}
	\bibinfo{author}{\bibfnamefont{L.}~\bibnamefont{Vitos}},
	\bibinfo{author}{\bibfnamefont{P.~A.} \bibnamefont{Korzhavyi}},
	\bibnamefont{and}
	\bibinfo{author}{\bibfnamefont{B.}~\bibnamefont{Johansson}},
	\bibinfo{journal}{Phys. Rev. Lett.} \textbf{\bibinfo{volume}{96}},
	\bibinfo{pages}{117210} (\bibinfo{year}{2006}).
	
	\bibitem[{\citenamefont{Sch{\"o}necker
			et~al.}(2013)\citenamefont{Sch{\"o}necker, Kwon, Johansson, and
			Vitos}}]{Schonecker2013}
	\bibinfo{author}{\bibfnamefont{S.}~\bibnamefont{Sch{\"o}necker}},
	\bibinfo{author}{\bibfnamefont{S.~K.} \bibnamefont{Kwon}},
	\bibinfo{author}{\bibfnamefont{B.}~\bibnamefont{Johansson}},
	\bibnamefont{and} \bibinfo{author}{\bibfnamefont{L.}~\bibnamefont{Vitos}},
	\bibinfo{journal}{J. Phys.: Condens. Matter} \textbf{\bibinfo{volume}{25}},
	\bibinfo{pages}{305002} (\bibinfo{year}{2013}).
	
	\bibitem[{\citenamefont{Ropo et~al.}(2005)\citenamefont{Ropo, Kokko, Vitos, and
			Koll{\'a}r}}]{Ropo2005}
	\bibinfo{author}{\bibfnamefont{M.}~\bibnamefont{Ropo}},
	\bibinfo{author}{\bibfnamefont{K.}~\bibnamefont{Kokko}},
	\bibinfo{author}{\bibfnamefont{L.}~\bibnamefont{Vitos}}, \bibnamefont{and}
	\bibinfo{author}{\bibfnamefont{J.}~\bibnamefont{Koll{\'a}r}},
	\bibinfo{journal}{Phys. Rev. B} \textbf{\bibinfo{volume}{71}},
	\bibinfo{pages}{045411} (\bibinfo{year}{2005}).
	
	\bibitem[{\citenamefont{Li et~al.}(2014{\natexlab{b}})\citenamefont{Li, Lu, Hu,
			Kwon, Johansson, and Vitos}}]{Li2014}
	\bibinfo{author}{\bibfnamefont{W.}~\bibnamefont{Li}},
	\bibinfo{author}{\bibfnamefont{S.}~\bibnamefont{Lu}},
	\bibinfo{author}{\bibfnamefont{Q.-M.} \bibnamefont{Hu}},
	\bibinfo{author}{\bibfnamefont{S.~K.} \bibnamefont{Kwon}},
	\bibinfo{author}{\bibfnamefont{B.}~\bibnamefont{Johansson}},
	\bibnamefont{and} \bibinfo{author}{\bibfnamefont{L.}~\bibnamefont{Vitos}},
	\bibinfo{journal}{J. Phys.: Condens. Matter} \textbf{\bibinfo{volume}{26}},
	\bibinfo{pages}{265005} (\bibinfo{year}{2014}{\natexlab{b}}).
	
	\bibitem[{\citenamefont{Lu et~al.}(2011)\citenamefont{Lu, Hu, Johansson, and
			Vitos}}]{Lu2011}
	\bibinfo{author}{\bibfnamefont{S.}~\bibnamefont{Lu}},
	\bibinfo{author}{\bibfnamefont{Q.-M.} \bibnamefont{Hu}},
	\bibinfo{author}{\bibfnamefont{B.}~\bibnamefont{Johansson}},
	\bibnamefont{and} \bibinfo{author}{\bibfnamefont{L.}~\bibnamefont{Vitos}},
	\bibinfo{journal}{Acta Mater.} \textbf{\bibinfo{volume}{59}},
	\bibinfo{pages}{5728 } (\bibinfo{year}{2011}).
	
	\bibitem[{\citenamefont{Lu et~al.}(2012)\citenamefont{Lu, Hu, Delczeg-Czirjak,
			Johansson, and Vitos}}]{Lu2012}
	\bibinfo{author}{\bibfnamefont{S.}~\bibnamefont{Lu}},
	\bibinfo{author}{\bibfnamefont{Q.-M.} \bibnamefont{Hu}},
	\bibinfo{author}{\bibfnamefont{E.~K.} \bibnamefont{Delczeg-Czirjak}},
	\bibinfo{author}{\bibfnamefont{B.}~\bibnamefont{Johansson}},
	\bibnamefont{and} \bibinfo{author}{\bibfnamefont{L.}~\bibnamefont{Vitos}},
	\bibinfo{journal}{Acta Mater.} \textbf{\bibinfo{volume}{60}},
	\bibinfo{pages}{4506 } (\bibinfo{year}{2012}).
	
	\bibitem[{\citenamefont{Perdew et~al.}(1996)\citenamefont{Perdew, Burke, and
			Ernzerhof}}]{Perdew1996}
	\bibinfo{author}{\bibfnamefont{J.~P.} \bibnamefont{Perdew}},
	\bibinfo{author}{\bibfnamefont{K.}~\bibnamefont{Burke}}, \bibnamefont{and}
	\bibinfo{author}{\bibfnamefont{M.}~\bibnamefont{Ernzerhof}},
	\bibinfo{journal}{Phys. Rev. Lett.} \textbf{\bibinfo{volume}{77}},
	\bibinfo{pages}{3865} (\bibinfo{year}{1996}).
	
	\bibitem[{\citenamefont{Zhang et~al.}(2010{\natexlab{a}})\citenamefont{Zhang,
			Punkkinen, Johansson, Hertzman, and Vitos}}]{Zhang2010}
	\bibinfo{author}{\bibfnamefont{H.}~\bibnamefont{Zhang}},
	\bibinfo{author}{\bibfnamefont{M.~P.} \bibnamefont{Punkkinen}},
	\bibinfo{author}{\bibfnamefont{B.}~\bibnamefont{Johansson}},
	\bibinfo{author}{\bibfnamefont{S.}~\bibnamefont{Hertzman}}, \bibnamefont{and}
	\bibinfo{author}{\bibfnamefont{L.}~\bibnamefont{Vitos}},
	\bibinfo{journal}{Phys. Rev. B} \textbf{\bibinfo{volume}{81}},
	\bibinfo{pages}{184105} (\bibinfo{year}{2010}{\natexlab{a}}).
	
	\bibitem[{\citenamefont{Asker et~al.}(2009)\citenamefont{Asker, Vitos, and
			Abrikosov}}]{Asker2009}
	\bibinfo{author}{\bibfnamefont{C.}~\bibnamefont{Asker}},
	\bibinfo{author}{\bibfnamefont{L.}~\bibnamefont{Vitos}}, \bibnamefont{and}
	\bibinfo{author}{\bibfnamefont{I.~A.} \bibnamefont{Abrikosov}},
	\bibinfo{journal}{Phys. Rev. B} \textbf{\bibinfo{volume}{79}},
	\bibinfo{pages}{214112} (\bibinfo{year}{2009}).
	
	\bibitem[{\citenamefont{Pitk{\"a}nen et~al.}(2009)\citenamefont{Pitk{\"a}nen,
			Alatalo, Puisto, Ropo, Kokko, Punkkinen, Olsson, Johansson, Hertzman, and
			Vitos}}]{Pitkanen2009}
	\bibinfo{author}{\bibfnamefont{H.}~\bibnamefont{Pitk{\"a}nen}},
	\bibinfo{author}{\bibfnamefont{M.}~\bibnamefont{Alatalo}},
	\bibinfo{author}{\bibfnamefont{A.}~\bibnamefont{Puisto}},
	\bibinfo{author}{\bibfnamefont{M.}~\bibnamefont{Ropo}},
	\bibinfo{author}{\bibfnamefont{K.}~\bibnamefont{Kokko}},
	\bibinfo{author}{\bibfnamefont{M.}~\bibnamefont{Punkkinen}},
	\bibinfo{author}{\bibfnamefont{P.}~\bibnamefont{Olsson}},
	\bibinfo{author}{\bibfnamefont{B.}~\bibnamefont{Johansson}},
	\bibinfo{author}{\bibfnamefont{S.}~\bibnamefont{Hertzman}}, \bibnamefont{and}
	\bibinfo{author}{\bibfnamefont{L.}~\bibnamefont{Vitos}},
	\bibinfo{journal}{Phys. Rev. B} \textbf{\bibinfo{volume}{79}},
	\bibinfo{pages}{024108} (\bibinfo{year}{2009}).
	
	\bibitem[{\citenamefont{Korzhavyi et~al.}(1995)\citenamefont{Korzhavyi, Ruban,
			Abrikosov, and Skriver}}]{Korzhavyi1995}
	\bibinfo{author}{\bibfnamefont{P.}~\bibnamefont{Korzhavyi}},
	\bibinfo{author}{\bibfnamefont{A.}~\bibnamefont{Ruban}},
	\bibinfo{author}{\bibfnamefont{I.}~\bibnamefont{Abrikosov}},
	\bibnamefont{and} \bibinfo{author}{\bibfnamefont{H.~L.}
		\bibnamefont{Skriver}}, \bibinfo{journal}{Phys. Rev. B}
	\textbf{\bibinfo{volume}{51}}, \bibinfo{pages}{5773} (\bibinfo{year}{1995}).
	
	\bibitem[{\citenamefont{Ruban and Skriver}(2002)}]{Ruban2002}
	\bibinfo{author}{\bibfnamefont{A.~V.} \bibnamefont{Ruban}} \bibnamefont{and}
	\bibinfo{author}{\bibfnamefont{H.~L.} \bibnamefont{Skriver}},
	\bibinfo{journal}{Phys. Rev. B} \textbf{\bibinfo{volume}{66}},
	\bibinfo{pages}{024201} (\bibinfo{year}{2002}).
	
	\bibitem[{\citenamefont{Wang et~al.}(2013)\citenamefont{Wang, Delczeg-Czirjak,
			Hu, Kokko, Johansson, and Vitos}}]{Guisheng2013}
	\bibinfo{author}{\bibfnamefont{G.-S.} \bibnamefont{Wang}},
	\bibinfo{author}{\bibfnamefont{E.~K.} \bibnamefont{Delczeg-Czirjak}},
	\bibinfo{author}{\bibfnamefont{Q.-M.} \bibnamefont{Hu}},
	\bibinfo{author}{\bibfnamefont{K.}~\bibnamefont{Kokko}},
	\bibinfo{author}{\bibfnamefont{B.}~\bibnamefont{Johansson}},
	\bibnamefont{and} \bibinfo{author}{\bibfnamefont{L.}~\bibnamefont{Vitos}},
	\bibinfo{journal}{J. Phys.: Condens. Matter} \textbf{\bibinfo{volume}{25}},
	\bibinfo{pages}{085401} (\bibinfo{year}{2013}).
	
	\bibitem[{\citenamefont{Moruzzi et~al.}(1988)\citenamefont{Moruzzi, Janak, and
			Schwarz}}]{Moruzzi1988}
	\bibinfo{author}{\bibfnamefont{V.}~\bibnamefont{Moruzzi}},
	\bibinfo{author}{\bibfnamefont{J.}~\bibnamefont{Janak}}, \bibnamefont{and}
	\bibinfo{author}{\bibfnamefont{K.}~\bibnamefont{Schwarz}},
	\bibinfo{journal}{Phys. Rev. B} \textbf{\bibinfo{volume}{37}},
	\bibinfo{pages}{790} (\bibinfo{year}{1988}).
	
	\bibitem[{\citenamefont{Zhang et~al.}(2010{\natexlab{b}})\citenamefont{Zhang,
			Lu, Punkkinen, Hu, Johansson, and Vitos}}]{Zhang2010a}
	\bibinfo{author}{\bibfnamefont{H.}~\bibnamefont{Zhang}},
	\bibinfo{author}{\bibfnamefont{S.}~\bibnamefont{Lu}},
	\bibinfo{author}{\bibfnamefont{M.~P.} \bibnamefont{Punkkinen}},
	\bibinfo{author}{\bibfnamefont{Q.-M.} \bibnamefont{Hu}},
	\bibinfo{author}{\bibfnamefont{B.}~\bibnamefont{Johansson}},
	\bibnamefont{and} \bibinfo{author}{\bibfnamefont{L.}~\bibnamefont{Vitos}},
	\bibinfo{journal}{Phys. Rev. B} \textbf{\bibinfo{volume}{82}},
	\bibinfo{pages}{132409} (\bibinfo{year}{2010}{\natexlab{b}}).
	
	\bibitem[{\citenamefont{Hill}(1952)}]{Hill1952}
	\bibinfo{author}{\bibfnamefont{R.}~\bibnamefont{Hill}}, \bibinfo{journal}{Proc.
		Phys. Sect. A} \textbf{\bibinfo{volume}{65}}, \bibinfo{pages}{349}
	(\bibinfo{year}{1952}).
	
	\bibitem[{\citenamefont{Fiorentini and Methfessel}(1996)}]{Fiorentini1996}
	\bibinfo{author}{\bibfnamefont{V.}~\bibnamefont{Fiorentini}} \bibnamefont{and}
	\bibinfo{author}{\bibfnamefont{M.}~\bibnamefont{Methfessel}},
	\bibinfo{journal}{J. Phys.: Condens. Matter} \textbf{\bibinfo{volume}{8}},
	\bibinfo{pages}{6525} (\bibinfo{year}{1996}).
	
	\bibitem[{\citenamefont{Punkkinen
			et~al.}(2011{\natexlab{a}})\citenamefont{Punkkinen, Hu, Kwon, Johansson,
			Koll{\'a}r, and Vitos}}]{Punkkinen2011}
	\bibinfo{author}{\bibfnamefont{M.~P.} \bibnamefont{Punkkinen}},
	\bibinfo{author}{\bibfnamefont{Q.-M.} \bibnamefont{Hu}},
	\bibinfo{author}{\bibfnamefont{S.~K.} \bibnamefont{Kwon}},
	\bibinfo{author}{\bibfnamefont{B.}~\bibnamefont{Johansson}},
	\bibinfo{author}{\bibfnamefont{J.}~\bibnamefont{Koll{\'a}r}},
	\bibnamefont{and} \bibinfo{author}{\bibfnamefont{L.}~\bibnamefont{Vitos}},
	\bibinfo{journal}{Philos. Mag.} \textbf{\bibinfo{volume}{91}},
	\bibinfo{pages}{3627} (\bibinfo{year}{2011}{\natexlab{a}}).
	
	\bibitem[{\citenamefont{Ruban et~al.}(1999)\citenamefont{Ruban, Skriver, and
			N{\o}rskov}}]{ruban1999surface}
	\bibinfo{author}{\bibfnamefont{A.}~\bibnamefont{Ruban}},
	\bibinfo{author}{\bibfnamefont{H.~L.} \bibnamefont{Skriver}},
	\bibnamefont{and} \bibinfo{author}{\bibfnamefont{J.~K.}
		\bibnamefont{N{\o}rskov}}, \bibinfo{journal}{Phys. Rev. B}
	\textbf{\bibinfo{volume}{59}}, \bibinfo{pages}{15990} (\bibinfo{year}{1999}).
	
	\bibitem[{\citenamefont{Caspersen et~al.}(2004)\citenamefont{Caspersen, Lew,
			Ortiz, and Carter}}]{Caspersen2004}
	\bibinfo{author}{\bibfnamefont{K.~J.} \bibnamefont{Caspersen}},
	\bibinfo{author}{\bibfnamefont{A.}~\bibnamefont{Lew}},
	\bibinfo{author}{\bibfnamefont{M.}~\bibnamefont{Ortiz}}, \bibnamefont{and}
	\bibinfo{author}{\bibfnamefont{E.~A.} \bibnamefont{Carter}},
	\bibinfo{journal}{Phys. Rev. Lett.} \textbf{\bibinfo{volume}{93}},
	\bibinfo{pages}{115501} (\bibinfo{year}{2004}).
	
	\bibitem[{\citenamefont{Vo{\v{c}}adlo et~al.}(1997)\citenamefont{Vo{\v{c}}adlo,
			de~Wijs, Kresse, Gillan, and Price}}]{Vovcadlo1997}
	\bibinfo{author}{\bibfnamefont{L.}~\bibnamefont{Vo{\v{c}}adlo}},
	\bibinfo{author}{\bibfnamefont{G.~A.} \bibnamefont{de~Wijs}},
	\bibinfo{author}{\bibfnamefont{G.}~\bibnamefont{Kresse}},
	\bibinfo{author}{\bibfnamefont{M.}~\bibnamefont{Gillan}}, \bibnamefont{and}
	\bibinfo{author}{\bibfnamefont{G.~D.} \bibnamefont{Price}},
	\bibinfo{journal}{Faraday Discuss.} \textbf{\bibinfo{volume}{106}},
	\bibinfo{pages}{205} (\bibinfo{year}{1997}).
	
	\bibitem[{\citenamefont{Rayne and Chandrasekhar}(1961)}]{Rayne1961}
	\bibinfo{author}{\bibfnamefont{J.}~\bibnamefont{Rayne}} \bibnamefont{and}
	\bibinfo{author}{\bibfnamefont{B.}~\bibnamefont{Chandrasekhar}},
	\bibinfo{journal}{Phys. Rev.} \textbf{\bibinfo{volume}{122}},
	\bibinfo{pages}{1714} (\bibinfo{year}{1961}).
	
	\bibitem[{\citenamefont{Mori et~al.}(2009)\citenamefont{Mori, Kimizuka, and
			Ogata}}]{Mori2009}
	\bibinfo{author}{\bibfnamefont{H.}~\bibnamefont{Mori}},
	\bibinfo{author}{\bibfnamefont{H.}~\bibnamefont{Kimizuka}}, \bibnamefont{and}
	\bibinfo{author}{\bibfnamefont{S.}~\bibnamefont{Ogata}}, \bibinfo{journal}{J.
		Japan Inst. Met.} \textbf{\bibinfo{volume}{73}}, \bibinfo{pages}{595}
	(\bibinfo{year}{2009}).
	
	\bibitem[{\citenamefont{Guo and Wang}(2000)}]{Guo2000}
	\bibinfo{author}{\bibfnamefont{G.}~\bibnamefont{Guo}} \bibnamefont{and}
	\bibinfo{author}{\bibfnamefont{H.}~\bibnamefont{Wang}},
	\bibinfo{journal}{Chin. J. Phys} \textbf{\bibinfo{volume}{38}},
	\bibinfo{pages}{949} (\bibinfo{year}{2000}).
	
	\bibitem[{\citenamefont{Sha and Cohen}(2006)}]{Sha2006}
	\bibinfo{author}{\bibfnamefont{X.}~\bibnamefont{Sha}} \bibnamefont{and}
	\bibinfo{author}{\bibfnamefont{R.}~\bibnamefont{Cohen}},
	\bibinfo{journal}{Phys. Rev. B} \textbf{\bibinfo{volume}{74}},
	\bibinfo{pages}{214111} (\bibinfo{year}{2006}).
	
	\bibitem[{\citenamefont{Tyson and Miller}(1977)}]{Tyson1977}
	\bibinfo{author}{\bibfnamefont{W.}~\bibnamefont{Tyson}} \bibnamefont{and}
	\bibinfo{author}{\bibfnamefont{W.}~\bibnamefont{Miller}},
	\bibinfo{journal}{Surf. Sci.} \textbf{\bibinfo{volume}{62}},
	\bibinfo{pages}{267} (\bibinfo{year}{1977}).
	
	\bibitem[{\citenamefont{Sutton and Hume-Rothery}(1955)}]{Sutton1955}
	\bibinfo{author}{\bibfnamefont{A.}~\bibnamefont{Sutton}} \bibnamefont{and}
	\bibinfo{author}{\bibfnamefont{W.}~\bibnamefont{Hume-Rothery}},
	\bibinfo{journal}{Philos. Mag.} \textbf{\bibinfo{volume}{46}},
	\bibinfo{pages}{1295} (\bibinfo{year}{1955}).
	
	\bibitem[{\citenamefont{Owen et~al.}(1937)\citenamefont{Owen, Yates, and
			Sully}}]{Owen1937}
	\bibinfo{author}{\bibfnamefont{E.}~\bibnamefont{Owen}},
	\bibinfo{author}{\bibfnamefont{E.}~\bibnamefont{Yates}}, \bibnamefont{and}
	\bibinfo{author}{\bibfnamefont{A.}~\bibnamefont{Sully}},
	\bibinfo{journal}{Proc. Phys. Soc.} \textbf{\bibinfo{volume}{49}},
	\bibinfo{pages}{307} (\bibinfo{year}{1937}).
	
	\bibitem[{\citenamefont{Ropo et~al.}(2008)\citenamefont{Ropo, Kokko, and
			Vitos}}]{Ropo2008}
	\bibinfo{author}{\bibfnamefont{M.}~\bibnamefont{Ropo}},
	\bibinfo{author}{\bibfnamefont{K.}~\bibnamefont{Kokko}}, \bibnamefont{and}
	\bibinfo{author}{\bibfnamefont{L.}~\bibnamefont{Vitos}},
	\bibinfo{journal}{Phys. Rev. B} \textbf{\bibinfo{volume}{77}},
	\bibinfo{pages}{195445} (\bibinfo{year}{2008}).
	
	\bibitem[{\citenamefont{Kulikov and Demangeat}(1997)}]{Kulikov1997}
	\bibinfo{author}{\bibfnamefont{N.}~\bibnamefont{Kulikov}} \bibnamefont{and}
	\bibinfo{author}{\bibfnamefont{C.}~\bibnamefont{Demangeat}},
	\bibinfo{journal}{Phys. Rev. B} \textbf{\bibinfo{volume}{55}},
	\bibinfo{pages}{3533} (\bibinfo{year}{1997}).
	
	\bibitem[{\citenamefont{Punkkinen
			et~al.}(2011{\natexlab{b}})\citenamefont{Punkkinen, Kwon, Koll{\'a}r,
			Johansson, and Vitos}}]{Punkkinen2011a}
	\bibinfo{author}{\bibfnamefont{M.}~\bibnamefont{Punkkinen}},
	\bibinfo{author}{\bibfnamefont{S.}~\bibnamefont{Kwon}},
	\bibinfo{author}{\bibfnamefont{J.}~\bibnamefont{Koll{\'a}r}},
	\bibinfo{author}{\bibfnamefont{B.}~\bibnamefont{Johansson}},
	\bibnamefont{and} \bibinfo{author}{\bibfnamefont{L.}~\bibnamefont{Vitos}},
	\bibinfo{journal}{Phys. Rev. Lett.} \textbf{\bibinfo{volume}{106}},
	\bibinfo{pages}{057202} (\bibinfo{year}{2011}{\natexlab{b}}).
	
	\bibitem[{\citenamefont{Li et~al.}(2014{\natexlab{c}})\citenamefont{Li,
			Sch\"onecker, Zhao, Johansson, and Vitos}}]{xiaoqing2014}
	\bibinfo{author}{\bibfnamefont{X.}~\bibnamefont{Li}},
	\bibinfo{author}{\bibfnamefont{S.}~\bibnamefont{Sch\"onecker}},
	\bibinfo{author}{\bibfnamefont{J.}~\bibnamefont{Zhao}},
	\bibinfo{author}{\bibfnamefont{B.}~\bibnamefont{Johansson}},
	\bibnamefont{and} \bibinfo{author}{\bibfnamefont{L.}~\bibnamefont{Vitos}},
	\bibinfo{journal}{Phys. Rev. B} \textbf{\bibinfo{volume}{90}},
	\bibinfo{pages}{024201} (\bibinfo{year}{2014}{\natexlab{c}}).
	
	\bibitem[{\citenamefont{Speich et~al.}(1972)\citenamefont{Speich, Schwoeble,
			and Leslie}}]{Speich1972}
	\bibinfo{author}{\bibfnamefont{G.}~\bibnamefont{Speich}},
	\bibinfo{author}{\bibfnamefont{A.}~\bibnamefont{Schwoeble}},
	\bibnamefont{and} \bibinfo{author}{\bibfnamefont{W.~C.}
		\bibnamefont{Leslie}}, \bibinfo{journal}{Metall Trans.}
	\textbf{\bibinfo{volume}{3}}, \bibinfo{pages}{2031} (\bibinfo{year}{1972}).
	
\end{thebibliography}

\end{document}